\documentclass[%
reprint,prl,
amsmath,amssymb,
aps
]{revtex4-1}

\usepackage{graphicx}% Include figure files
\usepackage{dcolumn}% Align table columns on decimal point
\usepackage{bm}% bold math
\usepackage{braket}
\usepackage{mathtools}
\usepackage{bbm}

\usepackage{afterpage}

\newcommand\blankpage{%
    \null
    \thispagestyle{empty}%
    \addtocounter{page}{-1}%
    \newpage}

\usepackage{hyperref}

\hypersetup{
	colorlinks=true,
	linkcolor=blue,
	citecolor=blue,
	filecolor=magenta,
	urlcolor=cyan
}

\newcommand{\iu}{{\rm i}}
\let\Re\relax
\let\Im\relax
\DeclareMathOperator{\Tr}{Tr}
\DeclareMathOperator{\Re}{Re}
\DeclareMathOperator{\Im}{Im}

\graphicspath{{../}}

\begin{document}

\def\papertitle{{Tenfold Way for Quadratic Lindbladians}}
\def\authornames{Simon Lieu, Max McGinley, and Nigel R.~Cooper}
\def\tcm{T.C.M. Group, Cavendish Laboratory, University of Cambridge, JJ Thomson Avenue, Cambridge, CB3 0HE, U.K.}

\title{\papertitle}
%\title{Topological classification of quadratic Lindbladians}
\author{Simon Lieu}
\affiliation{\tcm}
\author{Max McGinley}
\affiliation{\tcm}
\author{Nigel R.~Cooper}
\affiliation{\tcm}

\date{\today}

\begin{abstract}
We uncover a topological classification applicable to open fermionic systems governed by a general class of Lindblad master equations. These `quadratic Lindbladians' can be captured by a non-Hermitian single-particle matrix which describes internal dynamics as well as system-environment coupling. We show that this matrix must belong to one of ten non-Hermitian Bernard-LeClair symmetry classes which reduce to the Altland-Zirnbauer classes in the closed limit. The Lindblad spectrum admits a topological classification, which we show results in gapless edge excitations with finite lifetimes. Unlike previous studies of purely Hamiltonian or purely dissipative evolution, these topological edge modes are unconnected to the form of the steady state. We provide one-dimensional examples where the addition of dissipators can either preserve or destroy the closed classification of a model, highlighting the sensitivity of topological properties to details of the system-environment coupling.

\end{abstract}

\maketitle

\textit{Introduction.---} Topological band theory was developed to predict and explain robust features in the electronic structure of insulators and superconductors close to their ground states \cite{hasan2010, zhang2011}. While these ideas have already found fundamental applications  in quantum metrology \cite{vonKlitzing1986} and quantum computation \cite{alicea2012}, there has been a recent effort to understand the role of topology in the dynamics of many-body systems in highly non-equilibrium environments \cite{ rudner2013,barnett2015, else2016,vonKeyserlingk2016,potter2016, roy2017,mcginley2018, mcginley2019}.

A growing body of literature has been dedicated to studying topological aspects of ``non-Hermitian Hamiltonians,'' which generate non-unitary time evolution in certain dissipative classical and quantum settings \cite{ lieu2018b, kawabata2018BL, zhou2018BL,liu2019}. While this versatile approach applies in various limits, it is insufficient to describe the full time evolution of a generic open quantum many-body system coupled to a bath. An open system is described by a (possibly mixed) density matrix $\rho$ which propagates irreversibly due to dissipative coupling with its environment. For suitably generic baths, $\rho$ is governed by the Liouville equation: $\iu \dot{\rho} = \mathcal{L}(\rho)$, where $\mathcal{L}$ is the ``Lindbladian" -- a  non-Hermitian superoperator that acts linearly on $\rho(t)$. While calculating the complex spectrum of the Lindbladian can always be viewed as a non-Hermitian eigenvalue problem, $\mathcal{L}$ possesses an inherent structure which further constrains the topological  signatures of open systems.

In this paper, we show that there exists a robust topological classification of the full complex spectrum of the Lindbladian, $\mathcal{L}$, for the case of a Markovian bath with linear fermionic dissipators. In this case, the Lindblad spectral problem reduces to solving for the eigenvalues of a non-Hermitian quadratic Fermi operator \cite{prosen2008,prosen2010}. An understanding of the symmetry properties of this operator allows us  to compute the set of topologically distinct Lindblad spectra, which exhibit properties that are stable against continuous deformations. In particular, we make use of the real-line gap topological classification of  Bernard-LeClair symmetry classes \cite{bernard2002}, recently uncovered by Kawabata \textit{et al} \cite{kawabata2018BL}. 

Surprisingly, we find that our classification -- which applies in the presence of both dissipation and coherent internal dynamics -- 
differs qualitatively from the two limiting cases that have previously been much studied, of purely Hamiltonian systems (Hermitian Lindbladian) \cite{hasan2010, zhang2011} and of purely dissipative systems (anti-Hermitian Lindbladian) \cite{diehl2011,bardyn2013, goldstein2018, goldstein2019}. 

\begin{figure}
	\includegraphics[scale=1]{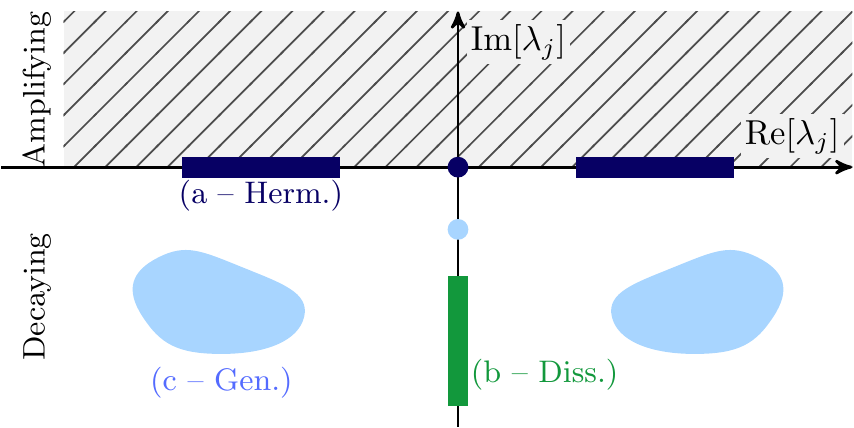}
	\caption{Complex spectra of one-dimensional examples of (a, dark blue) closed Hermitian systems; (b, green) purely dissipative systems studied in Refs.~\cite{diehl2011, bardyn2013, budich2015b}; and (c, light blue) generic quadratic Lindbladians studied in this work. Hermitian systems in a topological phase possess in-gap states with zero eigenvalue (dark blue dot), however the topology of purely dissipative systems is not reflected in the Lindblad spectrum. On the other hand, a quadratic Lindbladian which is gapped in the real direction can possess robust zero-frequency edge modes (light blue dot).}
	\label{fig:Spectrum}
\end{figure}

As in closed systems, the topological classification has consequences for dynamics near the system boundary. 
We show that a topologically non-trivial Lindbladian possesses robust edge modes whose phase-oscillation frequencies are pinned to lie in the energy gap, but which generically pick up finite lifetimes (See Fig.~\ref{fig:Spectrum}). (These edge modes will appear in spectroscopic measurements as broadened peaks within the bulk gap.)
However,  we find that, unlike previous classifications for purely Hamiltonian or purely dissipative systems, properties of the spectrum and steady state are completely independent: The existence of spectral edge modes implies nothing about the steady state density operator. 
For example, these universal topological properties of the complex excitation spectrum -- which have direct physical consequences in spectroscopy -- are unconnected to the classification of steady-state density matrices employed in Refs.~\cite{bardyn2013, viyuela2014, budich2015b, bardyn2018}. 
Our work highlights the various manifestations of band topology in a very general class of exactly solvable open systems, and provides formalisms which can be applied to understand generic interacting systems in future work.

\textit{Quadratic Lindbladians.---} Before discussing topological edge modes in an open environment, we describe the general setup considered in this work. %, i.e.~the definition of a quadratic Lindbladian.
Our starting point is the Gorini-Kossakowski-Sudarshan-Lindblad master equation
\begin{equation} \label{eq:lindMaster}
\iu  \frac{d\rho}{dt} =  \mathcal{L}(\rho) = [\mathcal{H}, \rho] + \iu \sum_\mu \left( 2 L_\mu \rho L_\mu^\dagger - \{L^\dagger_\mu L_\mu, \rho \}\right)
\end{equation}
which describes non-unitary time evolution of a density matrix $\rho$ subject to unitary dynamics generated by a Hamiltonian $\mathcal{H}$ and dissipation due to operators $L_\mu$ which can add and/or remove particles via a Markovian  %(memory-less)
environment \cite{lindblad1976}. Typically there exists a unique steady state $\rho_{SS}$ satisfying $\mathcal{L}\rho_{SS}=0$; all other eigenstates have complex eigenvalues with negative imaginary part, corresponding to terms decaying in time. Note that we have multiplied the typical definition of  $\mathcal{L}$ by $\iu$ such that the master equation resembles a non-Hermitian Schr\"{o}dinger equation: Real parts of eigenvalues (called energies) indicate phase oscillation frequencies of eigenstates, while negative imaginary parts correspond to the decay rate.

For a system of $N$ fermions, one can always solve for the spectrum $\lambda$ of the `Lindbladian' $\mathcal{L}$ by projecting onto some basis $\rho = \sum_{i,j} \rho_{i,j} \left| \phi_i \right\rangle \left\langle \phi_j \right| $, which has dimension $2^N \times 2^N = 2^{2N}$. Exact diagonalization of the resulting square matrix is  numerically expensive, since the basis grows exponentially with the number of particles. However, further progress can be made if the Hamiltonian is quadratic in Fermi operators, and the dissipators are linear -- such systems we refer to as \textit{quadratic Lindbladians}, and are the subject of this work. In this case, Prosen \cite{prosen2008,prosen2010} found that the spectrum of the Lindbladian can be found by diagonalizing a non-Hermitian fermionic superconductor with $2N$ particles in Bogoliubov-de Gennes  form. %This is the ``quadratic Lindbladian'' scenario considered in this work.
The factor of $2$ can be understood because we assign a fermion to both ``bra'' and ``ket'' space. The number of eigenstates is again $2^{2N}$ since each of the $2N$ Bogoliubov quasiparticles can either be excited or not.

%Once the Lindblad spectral problem is mapped to that of free fermions, it is natural to ask whether this non-Hermitian superconductor can host gapless topological edge modes; if so, what symmetries are important?

We briefly review this approach for $N$  complex fermions. The Hamiltonian and dissipators can be expressed in terms of $2N$ Majorana fermions
\begin{equation}
\mathcal{H} = \sum_{i,j=1}^{2N} \alpha_i H_{i,j} \alpha_j, \qquad L_\mu = \sum_{i=1}^{2N} l_{\mu, i} \alpha_i,
\label{eq:quadratic}
\end{equation}
where $H=H^\dagger, H = -H^T$. Majorana operators satisfy the anticommutation relation $\{\alpha_i,\alpha_j\}=2 \delta_{ij}$.
Define a $2N \times 2N$ Hermitian matrix $M = l^T l^*$. The Lindbladian can then be represented as a superoperator acting on a doubled Hilbert space spanned by $2N$ complex fermions $\{c_j\}$
\begin{equation} \label{eq:Lindblad1}
\mathcal{L} = 2
\left(\begin{array}{cc} \mathbf{c}^\dagger & \mathbf{c} \end{array}\right)
\left(\begin{array}{cc}  -Z^T & Y \\ 0 & Z   \end{array}\right)
\left(\begin{array}{c} \mathbf{c} \\ \mathbf{c}^\dagger \end{array}\right)
\end{equation}
where $\mathbf{c} = (c_1, \ldots, c_{2N})$, $Y=2 \Im[M]$, $Z=H+\iu \Re[M]$.  The $c$ superoperators explicitly act on the density matrix via: $c_j^\dagger \rho = (\alpha_j \rho +(\mathcal{P}^F \rho) \alpha_j)/2$ and $c_j \rho = (\alpha_j \rho -(\mathcal{P}^F \rho) \alpha_j)/2,$ where $\mathcal{P}^F$ is the fermion parity superoperator \cite{moos2019}. Due to this upper triangular form \eqref{eq:Lindblad1}, one can now diagonalize the Lindbladian in terms of $2N$ quasiparticles
\begin{equation} \label{eq:Lindblad2}
\mathcal{L} = 4 \sum_{j=1}^{2N} \lambda_j \bar{\beta}_j^\dagger \beta_j
\end{equation}
where $\lambda_j$ are the eigenvalues of the matrix $-Z$. Quasiparticles obey generalized fermionic statistics: $ \{ \bar{\beta}_i^{\dagger} , \beta_j \}= \delta_{i,j}, \, \{ \bar{\beta}_i^{\dagger} ,\bar{\beta}_j^{\dagger} \} = \{ \beta_i, \beta_j \} = 0$. In the doubled Hilbert space, the steady state is represented as a $2^{2N}$-dimensional vector that is annihilated by all quasiparticles: $\beta_i  \rho_{SS} =0$%$\beta_i  \left| \text{vac} \right\rangle =0$
. The states $\bar{\beta}_i^\dagger \rho_{SS}$ represent eigenoperators of $\mathcal{L}$, propagating with complex energy $4\lambda_i$. %$\bar{\beta}_i^\dagger  \left| \text{vac} \right\rangle$.

The single-particle Lindblad spectrum $\{ \lambda \}$ satisfies two generic conditions: (1) $\Im[\lambda_i] \leq 0$, since elements of the density matrix can only decay (not amplify) as a function of time, and  (2) Eigenvalues must come in anti-complex-conjugate pairs $\{ \lambda \}=\{- \lambda^* \} $ where the brackets indicate the set of spectral eigenvalues; this  ensures  Hermiticity of the density matrix at all times. 

\textit{Non-Hermitian tenfold way.---} In what follows, we will be interested in studying the robust features of the  complex Lindblad spectrum associated with a  topological insulator or superconductor in the presence of general linear fermionic dissipation. We begin by addressing the symmetries of the matrix whose eigenvalues determine the spectrum of  quadratic Lindbladians. From Eq.~\eqref{eq:Lindblad1}, the upper triangular structure of the matrix implies that the spectrum does not depend on $Y$, and hence  it is fully determined from the eigenvalues of the $2N$-dimensional square matrix $Z=H+\iu \Re[M]$.

The Hamiltonian of non-interacting fermions can be sorted into one of ten Altland-Zirnbauer \cite{altland1997} symmetry classes based the the presence or absence of the following three symmetries
\vspace{-2pt}\begin{subequations}
\label{eq:TFWHam}
\begin{align}
\text{TRS}:\qquad H & = U_T H^{*}U_T^{\dagger},& U_T U_T^{*}&=\pm\mathbb{I} \label{eq:trsHam}\\
\text{PHS}:\qquad H & = -U_C H^{*}U_C^{\dagger},& U_C U_C^{*}&=\pm\mathbb{I}\label{eq:phsHam}\\
\text{chiral}:\qquad H & = - U_S HU_S^{\dagger},& U_S^2&=\mathbb{I}\label{eq:chiralHam},
\end{align}
\end{subequations}
\vspace{-1pt}\noindent where the matrices $U_{T,C,S}$ are all unitary. Physically, these stem from time-reversal, particle-hole, and chiral (sublattice) symmetry respectively. Our use of Majorana fermions ensures that \eqref{eq:phsHam} is automatically satisfied with $U_C = \mathbb{I}$; however if charge is conserved then one can decouple particle and hole sectors, each of which separately does not respect PHS. A topological classification of non-interacting models based on these ten classes is called the tenfold way \cite{kitaev2009,ryu2010}, and describes symmetry-protected topological phases of free fermions. 

\begin{figure}
	\begin{centering}
		\includegraphics[scale=0.2]{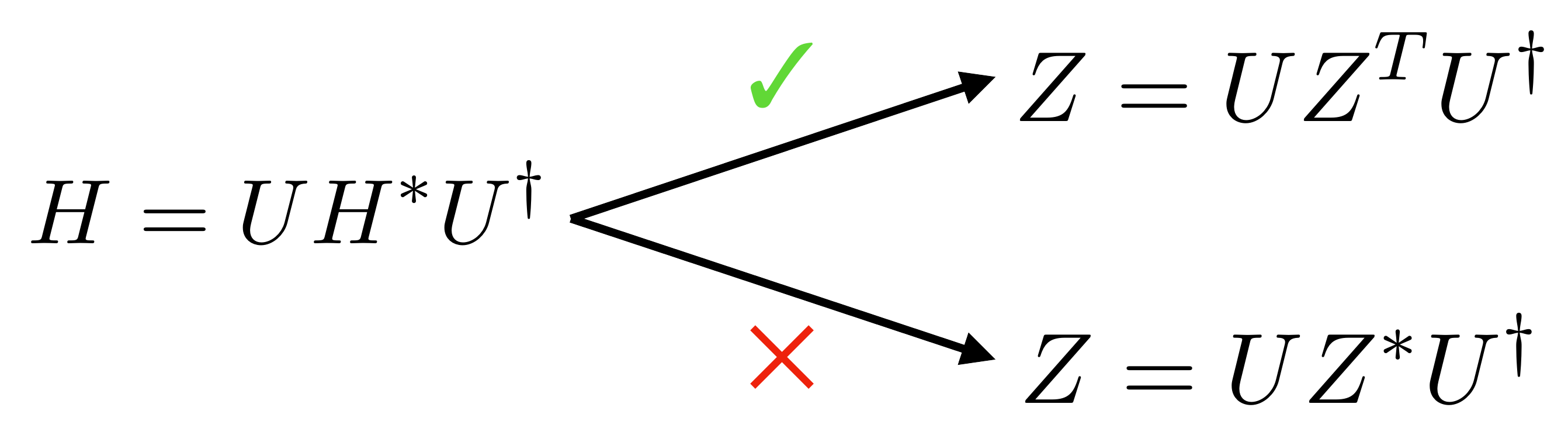}
		\par\end{centering}
	\caption{\label{fig:symSplit} Hermitian time-reversal symmetry (left) must be implemented using transposition rather than conjugation once non-Hermitian dissipative terms are included (right). ($Z=H+\iu \Re[M]$) }
\end{figure}

We  now ask whether $Z$ can inherit these symmetries once dissipators are introduced, i.e.~$ M \neq 0, Z \neq Z^\dagger$. If TRS is imposed on $Z$ in the form (\ref{eq:trsHam}), i.e. $Z = U_TZ^*U_T^\dagger$, then we will find that a damping mode with eigenvalue $\lambda$ must be paired with a mode of eigenvalue $\lambda^*$ -- this has the same frequency $\Re[\lambda]$ but a negative damping rate $\Im[\lambda]$, and is thus unphysical. (See Fig.~\ref{fig:symSplit}.) Similarly, PHS cannot be represented via an expression of the form $Z = - U_C Z^T U_C^\dagger,$ since this would ensure that eigenvalues come in positive-negative pairs: $\{ \lambda \} = \{ -\lambda \}$. Indeed $Z$ cannot respect any symmetry which associates a decaying mode with an amplifying one. We  find a unique way to extend the Hamiltonian symmetries \eqref{eq:TFWHam} to Lindbladian symmetries which does not suffer from this problem, namely
\vspace{-2pt}\begin{subequations}
	\label{eq:TFWZ}
\begin{align}
\text{TRS}:\qquad Z & = U_T Z^{T}U_T^{\dagger},& U_T U_T^{*}&=\pm\mathbb{I} \label{eq:trs}\\
\text{PHS}:\qquad Z & = -U_C Z^{*}U_C^{\dagger},& U_C U_C^{*}&=\pm\mathbb{I}\label{eq:phs}\\
\text{PAH}:\qquad Z & = -  U_S Z^{\dagger}U_S^{\dagger},& U_S^2&=\mathbb{I}\label{eq:chiral}.
\end{align}
\end{subequations}
\vspace{-1pt}\noindent Different combinations of these symmetries generate ten Lindbladian symmetry classes which reduce to the Altland-Zirnbauer classes in the absence of dissipation.  While the non-Hermitian Bernard-LeClair symmetries  generate a much larger number of unique  classes  compared to their Hermitian counterparts \cite{bernard2002}, the inherent structure of quadratic Lindbladians ensures that the spectral matrix $Z$ must belong to one of the ten classes defined above. Although the new form of time-reversal symmetry appears unusual, we show in the Supplementary Material \cite{SM} that this symmetry arises naturally when the microscopic Hamiltonian of the system and environment as a whole respect the Hermitian TRS \eqref{eq:trsHam} (even though the system alone propagates irreversably). Note also that pseudo-anti-Hermiticity (PAH) generalizes chiral symmetry, i.e.~it is guaranteed if a model has TRS and PHS.

Recent studies have used Bernard-LeClair symmetries to construct a topological classification for non-Hermitian models \cite{kawabata2018BL}. In this context, there exist different choices for defining a spectral gap -- some range of energy within which no bulk eigenvalues are present. The positivity condition $\Im[\lambda_i] \leq 0$ again puts constraints on these possibilities. If one chooses a point gap at the origin ($\lambda_i \neq 0$), or an imaginary line gap ($\Im[\lambda_i] \neq 0$), %{\bf *** imaginary***?}
 then the eigenvalues of $Z$ can be continuously deformed to a single point without crossing these gaps, and so an analysis under these conditions will not identify any robust spectral properties. However, one can choose a real line gap condition $\Re[\lambda_i] \neq 0$, i.e.~we insist that all bulk modes have a finite oscillation frequency [Fig.~\ref{fig:Spectrum}(c)]. Note that this is in stark contrast to the pure-dissipation case  \cite{diehl2011, bardyn2013, budich2015b}.

According to Ref.~\cite{kawabata2018BL}, the classification table for the ten BL classes which stem from Eqs.~\eqref{eq:TFWZ} under a real line gap is the same as that for the conventional tenfold way, once the non-Hermitian symmetry classes are associated with their corresponding Hermitian counterparts. The relevant bulk topological indices can be calculated for all the negative-frequency bands, and if their sum is non-zero then we expect in-gap states to appear at the system boundary, just as in Hermitian band theory. Since the gap is chosen along the imaginary axis, an edge mode of the Lindbladian will be pinned to zero frequency, but generically will have a finite damping rate, since the classification is only sensitive to $\Re[\lambda_i]$.

An intuitive picture is formed if one takes a topologically non-trivial system and gradually turns on dissipators without closing the frequency gap. If this procedure is carried out whilst at all times respecting the symmetries \eqref{eq:TFWZ}, then the topological classification of the new open system is identical to its closed precursor. The gapless edge modes of the Hermitian system will remain constrained to lie in the gap, and acquire a finite lifetime. Similarly, as was found for the SSH chain in Ref.~\cite{dangel2018}, topological invariants can be defined for the spectrum of the open system such that they are equal to those for the closed system. %This effect could be directly observed in spectroscopy measurements, which would

\textit{Independence of steady-state properties.---} In isolated systems, the topological properties of the ground state are reflected in the spectrum of the Hamiltonian. In open systems, the analogous state to consider is the non-equilibrium steady state $\mathcal{L}(\rho_{SS}) = 0$. Although $\rho_{SS}$ is generically not a pure state, one can still discuss its properties by using appropriate invariants for density matrices \cite{budich2015b}. Studies of systems with pure dissipation ($\mathcal{H} = 0$) have shown that an alternative tenfold way for open systems arises based on these properties \cite{diehl2011,bardyn2013, viyuela2014, bardyn2018}. One might expect that our spectral analysis reflects these steady state properties, in parallel with closed systems.

However, we find that the spectral and steady state topological properties of quadratic Lindbladians are independent. We prove this by showing that for any Lindbladian with a non-trivial steady state, there exists another Lindbladian with the same symmetries and spectrum, but with a trivial steady state. This auxiliary system has the same Hamiltonian, but the (generally complex) dissipators $\ell_{\mu, i}$  are replaced by real values $\tilde{\ell}_{\mu, i}$ which satisfy $\tilde{\ell}^T\tilde{\ell} = M \equiv \Re (\ell^T\ell^*)$. Because the matrix $Z$ depends only on $M$ and $H$, the spectrum is unaffected. However, one finds that $\rho_{SS} \propto \mathbbm{1}$, and is thus always a structureless ``trivial" steady state. In the Supplementary Material \cite{SM}, we show that a valid $\tilde{\ell}_{\mu, i}$ always exists and is sufficiently local such that one can define a continuous path of Lindbladians that leaves the spectrum invariant (e.g. without closing the gap in real frequency) yet connects the physical system to this auxiliary system with a trivial steady state. Hence, form of the  spectrum is unconnected to the form of the steady state.\\

Having uncovered the general symmetry-based topological classification of quadratic Lindbladians, we now illustrate its relevant features in the context of an example system.

%While previous studies have focused on generalizing ideas from ground state topology of a wavefunction to that of a steady-state density matrix, 
%study concrete examples in one dimension.  We demonstrate that the addition of dissipators can either preserve or destroy the equilibrium topological classification depending on whether or not they respect the symmetries of the subsystem model.

\textit{Dissipative Kitaev chain.---} We consider the Kitaev chain \cite{kitaev2001} in the presence of local, linear dissipators. The unitary evolution is generated via the Hamiltonian
\begin{equation} \label{eq:kitchain}
\mathcal{H}_{\text{Kit.}}= \iu \mu \sum_j  \alpha_{j,A} \alpha_{j,B}  + \iu \Delta \sum_j \alpha_{j,B} \alpha_{j+1,A}
\end{equation}
where $\alpha_{j,A/B}$ represent the two types of Majorana fermions on lattice site $j$ of $N$, and $\mu, \Delta \in \mathbb{R}$. We also consider $N-1$ dissipators which connect nearest-neighbor sites: $L_j=\gamma (\alpha_{j,A} + \iu \alpha_{j+1,B})$. A variant of this model has been studied previously \cite{moos2019}; however, we shall emphasize the importance of the non-Hermitian Bernard-LeClair symmetries which are responsible for the protection of gapless edge modes.

The Kitaev chain Hamiltonian falls into class BDI, which has a $\mathbb{Z}$ classification in 1D. In a Majorana basis, the first-quantized (matrix) Hamiltonian obeys the symmetries: $H=-H^*, H=\tau_z H^*\tau_z, H=-\tau_z H\tau_z$ where $\tau_z=\mathbb{I}_N \otimes \sigma_z$, and $\sigma_z$ is the Pauli matrix which acts on the Majorana sublattice index. If we turn on the dissipator strength $\gamma \neq 0$, then the dynamics of the open system is determined from the Lindblad spectrum, found explicitly by diagonalizing $Z$. $Z$ inherits the following symmetries:  $Z=-Z^*, Z=\tau_z Z^T\tau_z, Z=-\tau_z Z^\dagger \tau_z$. Indeed we find that such dissipators will keep the model in the same symmetry class, and we expect for edge modes to obey $\Re[\lambda_{\text{edge}}]=0$. For spinless fermions, any dissipator which can be written in the form: $L_\mu = e^{\iu\phi_\mu} \sum_j (\gamma_{\mu,j} \alpha_{j,A} + \iu \bar{\gamma}_{\mu,j} \alpha_{j,B}),$ for $\phi_\mu, \gamma_{\mu,j}, \bar{\gamma}_{\mu,j} \in \mathbb{R}$ will preserve the TRS condition \eqref{eq:trs}.

\begin{figure}
\begin{centering}
\includegraphics[scale=0.12]{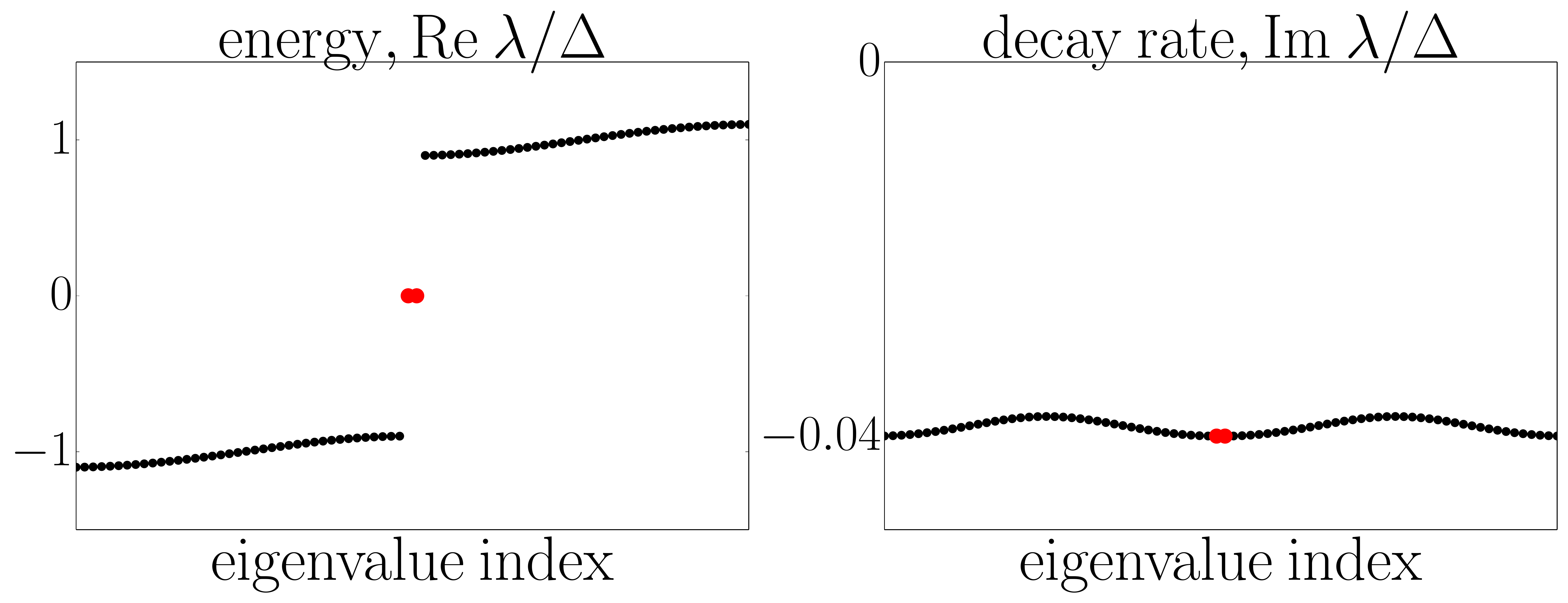}
\par\end{centering}
\caption{\label{fig:kitSpec} Lindblad spectrum  for the Kitaev chain with linear, nearest-neighbor dissipators, $\mu/\Delta=0.1, \gamma^2/\Delta=0.04$. A single edge mode exists on each side of the chain (red dots), and is symmetry-protected to obey $\Re[\lambda_{\text{edge}}]=0$. Majorana edge modes of the closed system can couple to fermionic dissipators and hence acquire a finite lifetime  $\Im[\lambda_{\text{edge}}]<0$.  }
\end{figure}

The spectrum is calculated numerically, and plotted in Fig.~\ref{fig:kitSpec}. We notice that indeed edge modes are constrained to obey $\Re[\lambda_{\text{edge}}]=0$, while the imaginary part of their energy becomes negative. Mathematically, this is due to pseudo-anti-Hermiticity: $Z=-\tau_z Z^\dagger \tau_z$ which implies $\lambda_{\text{edge}}=-\lambda_{\text{edge}}^* \implies \Re[\lambda_{\text{edge}}]=0, \Im[\lambda_{\text{edge}}]\neq 0 $ \cite{esaki2011}. We can also understand this behavior physically: The linear fermionic dissipators break fermion parity conservation of the closed Kitaev chain, hence Majorana modes at a given edge can couple to the environment and will acquire a finite lifetime (called quasiparticle poisoning) \cite{budich2012, carmele2015, moos2019}. If dissipators obeyed fermion parity then we would expect the steady state to retain its two-fold degeneracy due to decoupled parity sectors. (This type of dissipation falls outside the scope of quadratic Lindbladians.) Coupling to dissipators \textit{cannot}, however,  perturb the frequency of edge mode phase oscillations, since we have demonstrated that symmetries  protect these zero-frequency eigenstates of the Lindbladian.

The spectrum of the Lindbladian can be inferred from single-particle Green's functions in the frequency domain, i.e.~the Fourier transform of $\braket{\alpha_i(t) \alpha_j(0)}$. A particular eigenvalue $\lambda$ will give rise to a spectroscopic peak centred on $\Re [\lambda]$ with a characteristic width $\Im [\lambda]$. In experiment, these can be determined from linear response functions (see e.g.~Refs.~\cite{campos2016, albert2016}). For example, the
zero-bias tunneling peak characteristic of Majorana modes in topological superconductors should remain
centered at zero energy, but acquire a finite width.

In the Supplementary Material \cite{SM}, we discuss a different example (an open SSH chain) where the relevant symmetries can be either preserved or violated by the dissipators [whereas the PHS \eqref{eq:phs} intrinsic to superconducing systems cannot be broken].

\textit{Outlook.---} An immediate question is whether gapless edge modes can exist in the \textit{imaginary} spectrum, which would lead to robustly non-unique steady-state density matrices. While certain studies \cite{diehl2011, bardyn2013} have achieved this via ``topology by dissipation'' where Hamiltonian dynamics is fully switched off, such edge modes generically acquire a lifetime once Hamiltonian terms are added back, implying that this effect is fragile against such local perturbations. The existence of such a protected in-gap state for free fermions would require bands which amplify and bands which decay, such that the edge mode connects the two bulk bands. This scenario is forbidden, since the imaginary Lindblad spectrum is constrained to be non-positive.

%The topological signatures uncovered in this work are properties of the  Lindblad spectrum, and therefore do \textit{not} characterize the many-body steady state density matrix. While previous studies have focused on generalizing ideas from ground state topology of a wavefunction to that of a steady-state density matrix, our work points to topological signatures in the dynamics of an open many-body system as it approaches the steady state. Indeed we can show that topological invariants associated with the steady state are independent of the classification uncovered in this work: In the Supplementary Material \cite{SM}, we demonstrate that any quadratic Lindbladian can be smoothly deformed such that its excitation spectrum (as determined by  $Z$) remains constant while its steady-state evolves to the infinite-temperature density matrix (i.e.~the identity). This shows that the topology associated with $Z$ is independent of any non-trivial topology associated with the steady-state density matrix.

% paragraph on interactions
While we have limited our discussion to the case of  ``quadratic Lindbladians,'' we expect the topological edge modes described in this work to survive beyond this limit as non-Hermitian analogues of interacting symmetry-protected topological phases. For example, a quadratic Lindbladian respecting only PHS represents a dissipative topological superconductor, which will still be protected by fermion parity symmetry (as well as the Hermiticity-preserving nature of the Lindbladian) when solvability is broken. We also expect  that topological features of the spectrum and  steady state will remain decoupled in this limit: Unlike the Lindblad spectrum, the ground state of a closed system is not smoothly connected to the steady state of an open system with vanishingly small dissipation. Thus any topological properties of the former is not necessarily preserved in the latter. 

%Non-integrable Lindbladians include those with quadratic dissipators $L_\mu \sim  a_i a_j$, which can possess different symmetry properties to those accessible with linear dissipators, e.g.~$L_\mu$ could conserve the charge in the system.

%In the presence of quadratic dissipators $L_\mu$, the Lindblad superoperator \eqref{eq:lindMaster} resembles a non-Hermitian interacting (quartic) superconductor after doubling the Hilbert space to account for ``bra'' and ``ket'' degrees of freedom separately. A natural next step is to generalize the formalism of interacting symmetry-protected topological phases of fermions to include non-Hermitian terms. This would address the conditions required for the protection of gapless edge modes of the Lindbladian.  Notably, the Lindblad spectrum  always obeys: $\{\lambda\} = \{-\lambda^*\}$, required to  preserve Hermiticity of the density matrix at all times; this is a robust symmetry which can be utilized to protect zero-energy edge modes in \textit{any} open system.

We note in passing that the ten Lindblad symmetry classes uncovered in this paper may have interesting implications for the  spectral statistics of random dissipative systems \cite{can2019, karol2019, sa2019}. Imposing symmetries on the Lindbladian may result in universal  features of the complex spectrum, in analogy with the Altland-Zirnbauer random matrix classification of Hamiltonian dynamics.

In summary, we have discovered a topological classification which  constrains the dynamics of open femionic systems described by a Lindblad master equation. Specifically, we have demonstrated that  the addition of symmetry-preserving dissipators will ensure that edge modes of the Lindbladian have phase oscillations which are pinned to lie in the frequency gap, but will generically acquire a non-zero lifetime.  This causes the topological properties of the spectrum to decouple from those of the steady state. Our work provides a framework to systematically understand the protection of topological edge modes in the presence of both dissipation and internal dynamics.
 
 \begin{acknowledgments}
 	\emph{Acknowledgments.---}
 M.M.~thanks Jan Carl Budich for helpful discussions. This work was supported by the EPSRC and by a Simons Investigator Award. 

 \end{acknowledgments}

\bibliography{topoLind}

\begin{thebibliography}{46}%
\makeatletter
\providecommand \@ifxundefined [1]{%
 \@ifx{#1\undefined}
}%
\providecommand \@ifnum [1]{%
 \ifnum #1\expandafter \@firstoftwo
 \else \expandafter \@secondoftwo
 \fi
}%
\providecommand \@ifx [1]{%
 \ifx #1\expandafter \@firstoftwo
 \else \expandafter \@secondoftwo
 \fi
}%
\providecommand \natexlab [1]{#1}%
\providecommand \enquote  [1]{``#1''}%
\providecommand \bibnamefont  [1]{#1}%
\providecommand \bibfnamefont [1]{#1}%
\providecommand \citenamefont [1]{#1}%
\providecommand \href@noop [0]{\@secondoftwo}%
\providecommand \href [0]{\begingroup \@sanitize@url \@href}%
\providecommand \@href[1]{\@@startlink{#1}\@@href}%
\providecommand \@@href[1]{\endgroup#1\@@endlink}%
\providecommand \@sanitize@url [0]{\catcode `\\12\catcode `\$12\catcode
  `\&12\catcode `\#12\catcode `\^12\catcode `\_12\catcode `\%12\relax}%
\providecommand \@@startlink[1]{}%
\providecommand \@@endlink[0]{}%
\providecommand \url  [0]{\begingroup\@sanitize@url \@url }%
\providecommand \@url [1]{\endgroup\@href {#1}{\urlprefix }}%
\providecommand \urlprefix  [0]{URL }%
\providecommand \Eprint [0]{\href }%
\providecommand \doibase [0]{http://dx.doi.org/}%
\providecommand \selectlanguage [0]{\@gobble}%
\providecommand \bibinfo  [0]{\@secondoftwo}%
\providecommand \bibfield  [0]{\@secondoftwo}%
\providecommand \translation [1]{[#1]}%
\providecommand \BibitemOpen [0]{}%
\providecommand \bibitemStop [0]{}%
\providecommand \bibitemNoStop [0]{.\EOS\space}%
\providecommand \EOS [0]{\spacefactor3000\relax}%
\providecommand \BibitemShut  [1]{\csname bibitem#1\endcsname}%
\let\auto@bib@innerbib\@empty
%</preamble>
\bibitem [{\citenamefont {Hasan}\ and\ \citenamefont {Kane}(2010)}]{hasan2010}%
  \BibitemOpen
  \bibfield  {author} {\bibinfo {author} {\bibfnamefont {M.~Z.}\ \bibnamefont
  {Hasan}}\ and\ \bibinfo {author} {\bibfnamefont {C.~L.}\ \bibnamefont
  {Kane}},\ }\href {\doibase 10.1103/RevModPhys.82.3045} {\bibfield  {journal}
  {\bibinfo  {journal} {Rev. Mod. Phys.}\ }\textbf {\bibinfo {volume} {82}},\
  \bibinfo {pages} {3045} (\bibinfo {year} {2010})}\BibitemShut {NoStop}%
\bibitem [{\citenamefont {Qi}\ and\ \citenamefont {Zhang}(2011)}]{zhang2011}%
  \BibitemOpen
  \bibfield  {author} {\bibinfo {author} {\bibfnamefont {X.-L.}\ \bibnamefont
  {Qi}}\ and\ \bibinfo {author} {\bibfnamefont {S.-C.}\ \bibnamefont {Zhang}},\
  }\href {\doibase 10.1103/RevModPhys.83.1057} {\bibfield  {journal} {\bibinfo
  {journal} {Rev. Mod. Phys.}\ }\textbf {\bibinfo {volume} {83}},\ \bibinfo
  {pages} {1057} (\bibinfo {year} {2011})}\BibitemShut {NoStop}%
\bibitem [{\citenamefont {von Klitzing}(1986)}]{vonKlitzing1986}%
  \BibitemOpen
  \bibfield  {author} {\bibinfo {author} {\bibfnamefont {K.}~\bibnamefont {von
  Klitzing}},\ }\href {\doibase 10.1103/RevModPhys.58.519} {\bibfield
  {journal} {\bibinfo  {journal} {Rev. Mod. Phys.}\ }\textbf {\bibinfo {volume}
  {58}},\ \bibinfo {pages} {519} (\bibinfo {year} {1986})}\BibitemShut
  {NoStop}%
\bibitem [{\citenamefont {Alicea}(2012)}]{alicea2012}%
  \BibitemOpen
  \bibfield  {author} {\bibinfo {author} {\bibfnamefont {J.}~\bibnamefont
  {Alicea}},\ }\href {\doibase 10.1088/0034-4885/75/7/076501} {\bibfield
  {journal} {\bibinfo  {journal} {Reports on Progress in Physics}\ }\textbf
  {\bibinfo {volume} {75}},\ \bibinfo {pages} {076501} (\bibinfo {year}
  {2012})}\BibitemShut {NoStop}%
\bibitem [{\citenamefont {Rudner}\ \emph {et~al.}(2013)\citenamefont {Rudner},
  \citenamefont {Lindner}, \citenamefont {Berg},\ and\ \citenamefont
  {Levin}}]{rudner2013}%
  \BibitemOpen
  \bibfield  {author} {\bibinfo {author} {\bibfnamefont {M.~S.}\ \bibnamefont
  {Rudner}}, \bibinfo {author} {\bibfnamefont {N.~H.}\ \bibnamefont {Lindner}},
  \bibinfo {author} {\bibfnamefont {E.}~\bibnamefont {Berg}}, \ and\ \bibinfo
  {author} {\bibfnamefont {M.}~\bibnamefont {Levin}},\ }\href {\doibase
  10.1103/PhysRevX.3.031005} {\bibfield  {journal} {\bibinfo  {journal} {Phys.
  Rev. X}\ }\textbf {\bibinfo {volume} {3}},\ \bibinfo {pages} {031005}
  (\bibinfo {year} {2013})}\BibitemShut {NoStop}%
\bibitem [{\citenamefont {Galilo}\ \emph {et~al.}(2015)\citenamefont {Galilo},
  \citenamefont {Lee},\ and\ \citenamefont {Barnett}}]{barnett2015}%
  \BibitemOpen
  \bibfield  {author} {\bibinfo {author} {\bibfnamefont {B.}~\bibnamefont
  {Galilo}}, \bibinfo {author} {\bibfnamefont {D.~K.~K.}\ \bibnamefont {Lee}},
  \ and\ \bibinfo {author} {\bibfnamefont {R.}~\bibnamefont {Barnett}},\ }\href
  {\doibase 10.1103/PhysRevLett.115.245302} {\bibfield  {journal} {\bibinfo
  {journal} {Phys. Rev. Lett.}\ }\textbf {\bibinfo {volume} {115}},\ \bibinfo
  {pages} {245302} (\bibinfo {year} {2015})}\BibitemShut {NoStop}%
\bibitem [{\citenamefont {Else}\ and\ \citenamefont {Nayak}(2016)}]{else2016}%
  \BibitemOpen
  \bibfield  {author} {\bibinfo {author} {\bibfnamefont {D.~V.}\ \bibnamefont
  {Else}}\ and\ \bibinfo {author} {\bibfnamefont {C.}~\bibnamefont {Nayak}},\
  }\href {\doibase 10.1103/PhysRevB.93.201103} {\bibfield  {journal} {\bibinfo
  {journal} {Phys. Rev. B}\ }\textbf {\bibinfo {volume} {93}},\ \bibinfo
  {pages} {201103} (\bibinfo {year} {2016})}\BibitemShut {NoStop}%
\bibitem [{\citenamefont {von Keyserlingk}\ and\ \citenamefont
  {Sondhi}(2016)}]{vonKeyserlingk2016}%
  \BibitemOpen
  \bibfield  {author} {\bibinfo {author} {\bibfnamefont {C.~W.}\ \bibnamefont
  {von Keyserlingk}}\ and\ \bibinfo {author} {\bibfnamefont {S.~L.}\
  \bibnamefont {Sondhi}},\ }\href {\doibase 10.1103/PhysRevB.93.245145}
  {\bibfield  {journal} {\bibinfo  {journal} {Phys. Rev. B}\ }\textbf {\bibinfo
  {volume} {93}},\ \bibinfo {pages} {245145} (\bibinfo {year}
  {2016})}\BibitemShut {NoStop}%
\bibitem [{\citenamefont {Potter}\ \emph {et~al.}(2016)\citenamefont {Potter},
  \citenamefont {Morimoto},\ and\ \citenamefont {Vishwanath}}]{potter2016}%
  \BibitemOpen
  \bibfield  {author} {\bibinfo {author} {\bibfnamefont {A.~C.}\ \bibnamefont
  {Potter}}, \bibinfo {author} {\bibfnamefont {T.}~\bibnamefont {Morimoto}}, \
  and\ \bibinfo {author} {\bibfnamefont {A.}~\bibnamefont {Vishwanath}},\
  }\href {\doibase 10.1103/PhysRevX.6.041001} {\bibfield  {journal} {\bibinfo
  {journal} {Phys. Rev. X}\ }\textbf {\bibinfo {volume} {6}},\ \bibinfo {pages}
  {041001} (\bibinfo {year} {2016})}\BibitemShut {NoStop}%
\bibitem [{\citenamefont {Roy}\ and\ \citenamefont {Harper}(2017)}]{roy2017}%
  \BibitemOpen
  \bibfield  {author} {\bibinfo {author} {\bibfnamefont {R.}~\bibnamefont
  {Roy}}\ and\ \bibinfo {author} {\bibfnamefont {F.}~\bibnamefont {Harper}},\
  }\href {\doibase 10.1103/PhysRevB.96.155118} {\bibfield  {journal} {\bibinfo
  {journal} {Phys. Rev. B}\ }\textbf {\bibinfo {volume} {96}},\ \bibinfo
  {pages} {155118} (\bibinfo {year} {2017})}\BibitemShut {NoStop}%
\bibitem [{\citenamefont {McGinley}\ and\ \citenamefont
  {Cooper}(2018)}]{mcginley2018}%
  \BibitemOpen
  \bibfield  {author} {\bibinfo {author} {\bibfnamefont {M.}~\bibnamefont
  {McGinley}}\ and\ \bibinfo {author} {\bibfnamefont {N.~R.}\ \bibnamefont
  {Cooper}},\ }\href {\doibase 10.1103/PhysRevLett.121.090401} {\bibfield
  {journal} {\bibinfo  {journal} {Phys. Rev. Lett.}\ }\textbf {\bibinfo
  {volume} {121}},\ \bibinfo {pages} {090401} (\bibinfo {year}
  {2018})}\BibitemShut {NoStop}%
\bibitem [{\citenamefont {McGinley}\ and\ \citenamefont
  {Cooper}(2019)}]{mcginley2019}%
  \BibitemOpen
  \bibfield  {author} {\bibinfo {author} {\bibfnamefont {M.}~\bibnamefont
  {McGinley}}\ and\ \bibinfo {author} {\bibfnamefont {N.~R.}\ \bibnamefont
  {Cooper}},\ }\href {\doibase 10.1103/PhysRevB.99.075148} {\bibfield
  {journal} {\bibinfo  {journal} {Phys. Rev. B}\ }\textbf {\bibinfo {volume}
  {99}},\ \bibinfo {pages} {075148} (\bibinfo {year} {2019})}\BibitemShut
  {NoStop}%
\bibitem [{\citenamefont {Lieu}(2018)}]{lieu2018b}%
  \BibitemOpen
  \bibfield  {author} {\bibinfo {author} {\bibfnamefont {S.}~\bibnamefont
  {Lieu}},\ }\href {\doibase 10.1103/PhysRevB.98.115135} {\bibfield  {journal}
  {\bibinfo  {journal} {Phys. Rev. B}\ }\textbf {\bibinfo {volume} {98}},\
  \bibinfo {pages} {115135} (\bibinfo {year} {2018})}\BibitemShut {NoStop}%
\bibitem [{\citenamefont {Kawabata}\ \emph {et~al.}(2019)\citenamefont
  {Kawabata}, \citenamefont {Shiozaki}, \citenamefont {Ueda},\ and\
  \citenamefont {Sato}}]{kawabata2018BL}%
  \BibitemOpen
  \bibfield  {author} {\bibinfo {author} {\bibfnamefont {K.}~\bibnamefont
  {Kawabata}}, \bibinfo {author} {\bibfnamefont {K.}~\bibnamefont {Shiozaki}},
  \bibinfo {author} {\bibfnamefont {M.}~\bibnamefont {Ueda}}, \ and\ \bibinfo
  {author} {\bibfnamefont {M.}~\bibnamefont {Sato}},\ }\href {\doibase
  10.1103/PhysRevX.9.041015} {\bibfield  {journal} {\bibinfo  {journal} {Phys.
  Rev. X}\ }\textbf {\bibinfo {volume} {9}},\ \bibinfo {pages} {041015}
  (\bibinfo {year} {2019})}\BibitemShut {NoStop}%
\bibitem [{\citenamefont {Zhou}\ and\ \citenamefont {Lee}(2019)}]{zhou2018BL}%
  \BibitemOpen
  \bibfield  {author} {\bibinfo {author} {\bibfnamefont {H.}~\bibnamefont
  {Zhou}}\ and\ \bibinfo {author} {\bibfnamefont {J.~Y.}\ \bibnamefont {Lee}},\
  }\href {\doibase 10.1103/PhysRevB.99.235112} {\bibfield  {journal} {\bibinfo
  {journal} {Phys. Rev. B}\ }\textbf {\bibinfo {volume} {99}},\ \bibinfo
  {pages} {235112} (\bibinfo {year} {2019})}\BibitemShut {NoStop}%
\bibitem [{\citenamefont {Liu}\ and\ \citenamefont {Chen}(2019)}]{liu2019}%
  \BibitemOpen
  \bibfield  {author} {\bibinfo {author} {\bibfnamefont {C.-H.}\ \bibnamefont
  {Liu}}\ and\ \bibinfo {author} {\bibfnamefont {S.}~\bibnamefont {Chen}},\
  }\href {\doibase 10.1103/PhysRevB.100.144106} {\bibfield  {journal} {\bibinfo
   {journal} {Phys. Rev. B}\ }\textbf {\bibinfo {volume} {100}},\ \bibinfo
  {pages} {144106} (\bibinfo {year} {2019})}\BibitemShut {NoStop}%
\bibitem [{\citenamefont {Prosen}(2008)}]{prosen2008}%
  \BibitemOpen
  \bibfield  {author} {\bibinfo {author} {\bibfnamefont {T.}~\bibnamefont
  {Prosen}},\ }\href {\doibase 10.1088/1367-2630/10/4/043026} {\bibfield
  {journal} {\bibinfo  {journal} {New Journal of Physics}\ }\textbf {\bibinfo
  {volume} {10}},\ \bibinfo {pages} {043026} (\bibinfo {year}
  {2008})}\BibitemShut {NoStop}%
\bibitem [{\citenamefont {Prosen}(2010)}]{prosen2010}%
  \BibitemOpen
  \bibfield  {author} {\bibinfo {author} {\bibfnamefont {T.}~\bibnamefont
  {Prosen}},\ }\href {\doibase 10.1088/1742-5468/2010/07/p07020} {\bibfield
  {journal} {\bibinfo  {journal} {Journal of Statistical Mechanics: Theory and
  Experiment}\ }\textbf {\bibinfo {volume} {2010}},\ \bibinfo {pages} {P07020}
  (\bibinfo {year} {2010})}\BibitemShut {NoStop}%
\bibitem [{\citenamefont {Bernard}\ and\ \citenamefont
  {LeClair}(2002)}]{bernard2002}%
  \BibitemOpen
  \bibfield  {author} {\bibinfo {author} {\bibfnamefont {D.}~\bibnamefont
  {Bernard}}\ and\ \bibinfo {author} {\bibfnamefont {A.}~\bibnamefont
  {LeClair}},\ }\enquote {\bibinfo {title} {A classification of non-hermitian
  random matrices},}\ in\ \href@noop {} {\emph {\bibinfo {booktitle}
  {Statistical Field Theories}}},\ \bibinfo {editor} {edited by\ \bibinfo
  {editor} {\bibfnamefont {A.}~\bibnamefont {Cappelli}}\ and\ \bibinfo {editor}
  {\bibfnamefont {G.}~\bibnamefont {Mussardo}}}\ (\bibinfo  {publisher}
  {Springer Netherlands},\ \bibinfo {address} {Dordrecht},\ \bibinfo {year}
  {2002})\ pp.\ \bibinfo {pages} {207--214}\BibitemShut {NoStop}%
\bibitem [{\citenamefont {Diehl}\ \emph {et~al.}(2011)\citenamefont {Diehl},
  \citenamefont {Rico}, \citenamefont {Baranov},\ and\ \citenamefont
  {Zoller}}]{diehl2011}%
  \BibitemOpen
  \bibfield  {author} {\bibinfo {author} {\bibfnamefont {S.}~\bibnamefont
  {Diehl}}, \bibinfo {author} {\bibfnamefont {E.}~\bibnamefont {Rico}},
  \bibinfo {author} {\bibfnamefont {M.~A.}\ \bibnamefont {Baranov}}, \ and\
  \bibinfo {author} {\bibfnamefont {P.}~\bibnamefont {Zoller}},\ }\href
  {https://doi.org/10.1038/nphys2106} {\bibfield  {journal} {\bibinfo
  {journal} {Nature Physics}\ }\textbf {\bibinfo {volume} {7}},\ \bibinfo
  {pages} {971} (\bibinfo {year} {2011})}\BibitemShut {NoStop}%
\bibitem [{\citenamefont {Bardyn}\ \emph {et~al.}(2013)\citenamefont {Bardyn},
  \citenamefont {Baranov}, \citenamefont {Kraus}, \citenamefont {Rico},
  \citenamefont {{\.{I}}mamo{\u{g}}lu}, \citenamefont {Zoller},\ and\
  \citenamefont {Diehl}}]{bardyn2013}%
  \BibitemOpen
  \bibfield  {author} {\bibinfo {author} {\bibfnamefont {C.-E.}\ \bibnamefont
  {Bardyn}}, \bibinfo {author} {\bibfnamefont {M.~A.}\ \bibnamefont {Baranov}},
  \bibinfo {author} {\bibfnamefont {C.~V.}\ \bibnamefont {Kraus}}, \bibinfo
  {author} {\bibfnamefont {E.}~\bibnamefont {Rico}}, \bibinfo {author}
  {\bibfnamefont {A.}~\bibnamefont {{\.{I}}mamo{\u{g}}lu}}, \bibinfo {author}
  {\bibfnamefont {P.}~\bibnamefont {Zoller}}, \ and\ \bibinfo {author}
  {\bibfnamefont {S.}~\bibnamefont {Diehl}},\ }\href {\doibase
  10.1088/1367-2630/15/8/085001} {\bibfield  {journal} {\bibinfo  {journal}
  {New Journal of Physics}\ }\textbf {\bibinfo {volume} {15}},\ \bibinfo
  {pages} {085001} (\bibinfo {year} {2013})}\BibitemShut {NoStop}%
\bibitem [{\citenamefont {Goldstein}(2018)}]{goldstein2018}%
  \BibitemOpen
  \bibfield  {author} {\bibinfo {author} {\bibfnamefont {M.}~\bibnamefont
  {Goldstein}},\ }\href@noop {} {\bibfield  {journal} {\bibinfo  {journal}
  {arXiv:1810.12050}\ } (\bibinfo {year} {2018})}\BibitemShut {NoStop}%
\bibitem [{\citenamefont {Shavit}\ and\ \citenamefont
  {Goldstein}(2019)}]{goldstein2019}%
  \BibitemOpen
  \bibfield  {author} {\bibinfo {author} {\bibfnamefont {G.}~\bibnamefont
  {Shavit}}\ and\ \bibinfo {author} {\bibfnamefont {M.}~\bibnamefont
  {Goldstein}},\ }\href@noop {} {\bibfield  {journal} {\bibinfo  {journal}
  {arXiv:1903.05336}\ } (\bibinfo {year} {2019})}\BibitemShut {NoStop}%
\bibitem [{\citenamefont {Viyuela}\ \emph {et~al.}(2014)\citenamefont
  {Viyuela}, \citenamefont {Rivas},\ and\ \citenamefont
  {Martin-Delgado}}]{viyuela2014}%
  \BibitemOpen
  \bibfield  {author} {\bibinfo {author} {\bibfnamefont {O.}~\bibnamefont
  {Viyuela}}, \bibinfo {author} {\bibfnamefont {A.}~\bibnamefont {Rivas}}, \
  and\ \bibinfo {author} {\bibfnamefont {M.~A.}\ \bibnamefont
  {Martin-Delgado}},\ }\href {\doibase 10.1103/PhysRevLett.112.130401}
  {\bibfield  {journal} {\bibinfo  {journal} {Phys. Rev. Lett.}\ }\textbf
  {\bibinfo {volume} {112}},\ \bibinfo {pages} {130401} (\bibinfo {year}
  {2014})}\BibitemShut {NoStop}%
\bibitem [{\citenamefont {Budich}\ and\ \citenamefont
  {Diehl}(2015)}]{budich2015b}%
  \BibitemOpen
  \bibfield  {author} {\bibinfo {author} {\bibfnamefont {J.~C.}\ \bibnamefont
  {Budich}}\ and\ \bibinfo {author} {\bibfnamefont {S.}~\bibnamefont {Diehl}},\
  }\href {\doibase 10.1103/PhysRevB.91.165140} {\bibfield  {journal} {\bibinfo
  {journal} {Phys. Rev. B}\ }\textbf {\bibinfo {volume} {91}},\ \bibinfo
  {pages} {165140} (\bibinfo {year} {2015})}\BibitemShut {NoStop}%
\bibitem [{\citenamefont {Bardyn}\ \emph {et~al.}(2018)\citenamefont {Bardyn},
  \citenamefont {Wawer}, \citenamefont {Altland}, \citenamefont
  {Fleischhauer},\ and\ \citenamefont {Diehl}}]{bardyn2018}%
  \BibitemOpen
  \bibfield  {author} {\bibinfo {author} {\bibfnamefont {C.-E.}\ \bibnamefont
  {Bardyn}}, \bibinfo {author} {\bibfnamefont {L.}~\bibnamefont {Wawer}},
  \bibinfo {author} {\bibfnamefont {A.}~\bibnamefont {Altland}}, \bibinfo
  {author} {\bibfnamefont {M.}~\bibnamefont {Fleischhauer}}, \ and\ \bibinfo
  {author} {\bibfnamefont {S.}~\bibnamefont {Diehl}},\ }\href {\doibase
  10.1103/PhysRevX.8.011035} {\bibfield  {journal} {\bibinfo  {journal} {Phys.
  Rev. X}\ }\textbf {\bibinfo {volume} {8}},\ \bibinfo {pages} {011035}
  (\bibinfo {year} {2018})}\BibitemShut {NoStop}%
\bibitem [{\citenamefont {Lindblad}(1976)}]{lindblad1976}%
  \BibitemOpen
  \bibfield  {author} {\bibinfo {author} {\bibfnamefont {G.}~\bibnamefont
  {Lindblad}},\ }\href {https://link.springer.com/article/10.1007/BF01608499}
  {\bibfield  {journal} {\bibinfo  {journal} {Comm.~Math.~Phys.}\ }\textbf
  {\bibinfo {volume} {48}},\ \bibinfo {pages} {119} (\bibinfo {year}
  {1976})}\BibitemShut {NoStop}%
\bibitem [{\citenamefont {van Caspel}\ \emph {et~al.}(2019)\citenamefont {van
  Caspel}, \citenamefont {Arze},\ and\ \citenamefont {Castillo}}]{moos2019}%
  \BibitemOpen
  \bibfield  {author} {\bibinfo {author} {\bibfnamefont {M.}~\bibnamefont {van
  Caspel}}, \bibinfo {author} {\bibfnamefont {S.~E.~T.}\ \bibnamefont {Arze}},
  \ and\ \bibinfo {author} {\bibfnamefont {I.~P.}\ \bibnamefont {Castillo}},\
  }\href {\doibase 10.21468/SciPostPhys.6.2.026} {\bibfield  {journal}
  {\bibinfo  {journal} {SciPost Phys.}\ }\textbf {\bibinfo {volume} {6}},\
  \bibinfo {pages} {26} (\bibinfo {year} {2019})}\BibitemShut {NoStop}%
\bibitem [{\citenamefont {Altland}\ and\ \citenamefont
  {Zirnbauer}(1997)}]{altland1997}%
  \BibitemOpen
  \bibfield  {author} {\bibinfo {author} {\bibfnamefont {A.}~\bibnamefont
  {Altland}}\ and\ \bibinfo {author} {\bibfnamefont {M.~R.}\ \bibnamefont
  {Zirnbauer}},\ }\href {\doibase 10.1103/PhysRevB.55.1142} {\bibfield
  {journal} {\bibinfo  {journal} {Phys. Rev. B}\ }\textbf {\bibinfo {volume}
  {55}},\ \bibinfo {pages} {1142} (\bibinfo {year} {1997})}\BibitemShut
  {NoStop}%
\bibitem [{\citenamefont {Kitaev}(2009)}]{kitaev2009}%
  \BibitemOpen
  \bibfield  {author} {\bibinfo {author} {\bibfnamefont {A.}~\bibnamefont
  {Kitaev}},\ }\href {\doibase 10.1063/1.3149495} {\bibfield  {journal}
  {\bibinfo  {journal} {AIP Conference Proceedings}\ }\textbf {\bibinfo
  {volume} {1134}},\ \bibinfo {pages} {22} (\bibinfo {year}
  {2009})}\BibitemShut {NoStop}%
\bibitem [{\citenamefont {Ryu}\ \emph {et~al.}(2010)\citenamefont {Ryu},
  \citenamefont {Schnyder}, \citenamefont {Furusaki},\ and\ \citenamefont
  {Ludwig}}]{ryu2010}%
  \BibitemOpen
  \bibfield  {author} {\bibinfo {author} {\bibfnamefont {S.}~\bibnamefont
  {Ryu}}, \bibinfo {author} {\bibfnamefont {A.~P.}\ \bibnamefont {Schnyder}},
  \bibinfo {author} {\bibfnamefont {A.}~\bibnamefont {Furusaki}}, \ and\
  \bibinfo {author} {\bibfnamefont {A.~W.~W.}\ \bibnamefont {Ludwig}},\ }\href
  {http://stacks.iop.org/1367-2630/12/i=6/a=065010} {\bibfield  {journal}
  {\bibinfo  {journal} {New Journal of Physics}\ }\textbf {\bibinfo {volume}
  {12}},\ \bibinfo {pages} {065010} (\bibinfo {year} {2010})}\BibitemShut
  {NoStop}%
\bibitem [{SM()}]{SM}%
  \BibitemOpen
  \href@noop {} {}\bibinfo {note} {See the Supplementary Material for numerical
  results on the dissipative SSH chain, a discussion of the conditions to
  satisfy the non-Hermitian time-reversal symmetry \eqref{eq:trs}, and a proof
  that spectral and steady-state properties of the Lindbladian are indepedent.
  Contains Refs.~\cite{breuer2002,chiu2016,kitaev2006,su1979}}\BibitemShut
  {NoStop}%
\bibitem [{\citenamefont {Dangel}\ \emph {et~al.}(2018)\citenamefont {Dangel},
  \citenamefont {Wagner}, \citenamefont {Cartarius}, \citenamefont {Main},\
  and\ \citenamefont {Wunner}}]{dangel2018}%
  \BibitemOpen
  \bibfield  {author} {\bibinfo {author} {\bibfnamefont {F.}~\bibnamefont
  {Dangel}}, \bibinfo {author} {\bibfnamefont {M.}~\bibnamefont {Wagner}},
  \bibinfo {author} {\bibfnamefont {H.}~\bibnamefont {Cartarius}}, \bibinfo
  {author} {\bibfnamefont {J.}~\bibnamefont {Main}}, \ and\ \bibinfo {author}
  {\bibfnamefont {G.}~\bibnamefont {Wunner}},\ }\href {\doibase
  10.1103/PhysRevA.98.013628} {\bibfield  {journal} {\bibinfo  {journal} {Phys.
  Rev. A}\ }\textbf {\bibinfo {volume} {98}},\ \bibinfo {pages} {013628}
  (\bibinfo {year} {2018})}\BibitemShut {NoStop}%
\bibitem [{\citenamefont {Kitaev}(2001)}]{kitaev2001}%
  \BibitemOpen
  \bibfield  {author} {\bibinfo {author} {\bibfnamefont {A.~Y.}\ \bibnamefont
  {Kitaev}},\ }\href {http://stacks.iop.org/1063-7869/44/i=10S/a=S29}
  {\bibfield  {journal} {\bibinfo  {journal} {Physics-Uspekhi}\ }\textbf
  {\bibinfo {volume} {44}},\ \bibinfo {pages} {131} (\bibinfo {year}
  {2001})}\BibitemShut {NoStop}%
\bibitem [{\citenamefont {Esaki}\ \emph {et~al.}(2011)\citenamefont {Esaki},
  \citenamefont {Sato}, \citenamefont {Hasebe},\ and\ \citenamefont
  {Kohmoto}}]{esaki2011}%
  \BibitemOpen
  \bibfield  {author} {\bibinfo {author} {\bibfnamefont {K.}~\bibnamefont
  {Esaki}}, \bibinfo {author} {\bibfnamefont {M.}~\bibnamefont {Sato}},
  \bibinfo {author} {\bibfnamefont {K.}~\bibnamefont {Hasebe}}, \ and\ \bibinfo
  {author} {\bibfnamefont {M.}~\bibnamefont {Kohmoto}},\ }\href {\doibase
  10.1103/PhysRevB.84.205128} {\bibfield  {journal} {\bibinfo  {journal} {Phys.
  Rev. B}\ }\textbf {\bibinfo {volume} {84}},\ \bibinfo {pages} {205128}
  (\bibinfo {year} {2011})}\BibitemShut {NoStop}%
\bibitem [{\citenamefont {Budich}\ \emph {et~al.}(2012)\citenamefont {Budich},
  \citenamefont {Walter},\ and\ \citenamefont {Trauzettel}}]{budich2012}%
  \BibitemOpen
  \bibfield  {author} {\bibinfo {author} {\bibfnamefont {J.~C.}\ \bibnamefont
  {Budich}}, \bibinfo {author} {\bibfnamefont {S.}~\bibnamefont {Walter}}, \
  and\ \bibinfo {author} {\bibfnamefont {B.}~\bibnamefont {Trauzettel}},\
  }\href {\doibase 10.1103/PhysRevB.85.121405} {\bibfield  {journal} {\bibinfo
  {journal} {Phys. Rev. B}\ }\textbf {\bibinfo {volume} {85}},\ \bibinfo
  {pages} {121405} (\bibinfo {year} {2012})}\BibitemShut {NoStop}%
\bibitem [{\citenamefont {Carmele}\ \emph {et~al.}(2015)\citenamefont
  {Carmele}, \citenamefont {Heyl}, \citenamefont {Kraus},\ and\ \citenamefont
  {Dalmonte}}]{carmele2015}%
  \BibitemOpen
  \bibfield  {author} {\bibinfo {author} {\bibfnamefont {A.}~\bibnamefont
  {Carmele}}, \bibinfo {author} {\bibfnamefont {M.}~\bibnamefont {Heyl}},
  \bibinfo {author} {\bibfnamefont {C.}~\bibnamefont {Kraus}}, \ and\ \bibinfo
  {author} {\bibfnamefont {M.}~\bibnamefont {Dalmonte}},\ }\href {\doibase
  10.1103/PhysRevB.92.195107} {\bibfield  {journal} {\bibinfo  {journal} {Phys.
  Rev. B}\ }\textbf {\bibinfo {volume} {92}},\ \bibinfo {pages} {195107}
  (\bibinfo {year} {2015})}\BibitemShut {NoStop}%
\bibitem [{\citenamefont {Campos~Venuti}\ and\ \citenamefont
  {Zanardi}(2016)}]{campos2016}%
  \BibitemOpen
  \bibfield  {author} {\bibinfo {author} {\bibfnamefont {L.}~\bibnamefont
  {Campos~Venuti}}\ and\ \bibinfo {author} {\bibfnamefont {P.}~\bibnamefont
  {Zanardi}},\ }\href {\doibase 10.1103/PhysRevA.93.032101} {\bibfield
  {journal} {\bibinfo  {journal} {Phys. Rev. A}\ }\textbf {\bibinfo {volume}
  {93}},\ \bibinfo {pages} {032101} (\bibinfo {year} {2016})}\BibitemShut
  {NoStop}%
\bibitem [{\citenamefont {Albert}\ \emph {et~al.}(2016)\citenamefont {Albert},
  \citenamefont {Bradlyn}, \citenamefont {Fraas},\ and\ \citenamefont
  {Jiang}}]{albert2016}%
  \BibitemOpen
  \bibfield  {author} {\bibinfo {author} {\bibfnamefont {V.~V.}\ \bibnamefont
  {Albert}}, \bibinfo {author} {\bibfnamefont {B.}~\bibnamefont {Bradlyn}},
  \bibinfo {author} {\bibfnamefont {M.}~\bibnamefont {Fraas}}, \ and\ \bibinfo
  {author} {\bibfnamefont {L.}~\bibnamefont {Jiang}},\ }\href {\doibase
  10.1103/PhysRevX.6.041031} {\bibfield  {journal} {\bibinfo  {journal} {Phys.
  Rev. X}\ }\textbf {\bibinfo {volume} {6}},\ \bibinfo {pages} {041031}
  (\bibinfo {year} {2016})}\BibitemShut {NoStop}%
\bibitem [{\citenamefont {Can}\ \emph {et~al.}(2019)\citenamefont {Can},
  \citenamefont {Oganesyan}, \citenamefont {Orgad},\ and\ \citenamefont
  {Gopalakrishnan}}]{can2019}%
  \BibitemOpen
  \bibfield  {author} {\bibinfo {author} {\bibfnamefont {T.}~\bibnamefont
  {Can}}, \bibinfo {author} {\bibfnamefont {V.}~\bibnamefont {Oganesyan}},
  \bibinfo {author} {\bibfnamefont {D.}~\bibnamefont {Orgad}}, \ and\ \bibinfo
  {author} {\bibfnamefont {S.}~\bibnamefont {Gopalakrishnan}},\ }\href
  {\doibase 10.1103/PhysRevLett.123.234103} {\bibfield  {journal} {\bibinfo
  {journal} {Phys. Rev. Lett.}\ }\textbf {\bibinfo {volume} {123}},\ \bibinfo
  {pages} {234103} (\bibinfo {year} {2019})}\BibitemShut {NoStop}%
\bibitem [{\citenamefont {Denisov}\ \emph {et~al.}(2019)\citenamefont
  {Denisov}, \citenamefont {Laptyeva}, \citenamefont {Tarnowski}, \citenamefont
  {Chru\ifmmode \acute{s}\else \'{s}\fi{}ci\ifmmode~\acute{n}\else
  \'{n}\fi{}ski},\ and\ \citenamefont {\ifmmode~\dot{Z}\else
  \.{Z}\fi{}yczkowski}}]{karol2019}%
  \BibitemOpen
  \bibfield  {author} {\bibinfo {author} {\bibfnamefont {S.}~\bibnamefont
  {Denisov}}, \bibinfo {author} {\bibfnamefont {T.}~\bibnamefont {Laptyeva}},
  \bibinfo {author} {\bibfnamefont {W.}~\bibnamefont {Tarnowski}}, \bibinfo
  {author} {\bibfnamefont {D.}~\bibnamefont {Chru\ifmmode \acute{s}\else
  \'{s}\fi{}ci\ifmmode~\acute{n}\else \'{n}\fi{}ski}}, \ and\ \bibinfo {author}
  {\bibfnamefont {K.}~\bibnamefont {\ifmmode~\dot{Z}\else
  \.{Z}\fi{}yczkowski}},\ }\href {\doibase 10.1103/PhysRevLett.123.140403}
  {\bibfield  {journal} {\bibinfo  {journal} {Phys. Rev. Lett.}\ }\textbf
  {\bibinfo {volume} {123}},\ \bibinfo {pages} {140403} (\bibinfo {year}
  {2019})}\BibitemShut {NoStop}%
\bibitem [{\citenamefont {{S{\'a}}}\ \emph {et~al.}(2019)\citenamefont
  {{S{\'a}}}, \citenamefont {{Ribeiro}},\ and\ \citenamefont
  {{Prosen}}}]{sa2019}%
  \BibitemOpen
  \bibfield  {author} {\bibinfo {author} {\bibfnamefont {L.}~\bibnamefont
  {{S{\'a}}}}, \bibinfo {author} {\bibfnamefont {P.}~\bibnamefont {{Ribeiro}}},
  \ and\ \bibinfo {author} {\bibfnamefont {T.}~\bibnamefont {{Prosen}}},\
  }\href@noop {} {\ ,\ \bibinfo {eid} {arXiv:1905.02155} (\bibinfo {year}
  {2019})},\ \Eprint {http://arxiv.org/abs/1905.02155} {arXiv:1905.02155
  [quant-ph]} \BibitemShut {NoStop}%
\bibitem [{\citenamefont {Breuer}\ and\ \citenamefont
  {Petruccione}(2002)}]{breuer2002}%
  \BibitemOpen
  \bibfield  {author} {\bibinfo {author} {\bibfnamefont {H.}~\bibnamefont
  {Breuer}}\ and\ \bibinfo {author} {\bibfnamefont {F.}~\bibnamefont
  {Petruccione}},\ }\href {https://books.google.co.uk/books?id=w2UOnwEACAAJ}
  {\emph {\bibinfo {title} {The Theory of Open Quantum Systems}}}\ (\bibinfo
  {publisher} {Oxford University Press},\ \bibinfo {year} {2002})\BibitemShut
  {NoStop}%
\bibitem [{\citenamefont {Chiu}\ \emph {et~al.}(2016)\citenamefont {Chiu},
  \citenamefont {Teo}, \citenamefont {Schnyder},\ and\ \citenamefont
  {Ryu}}]{chiu2016}%
  \BibitemOpen
  \bibfield  {author} {\bibinfo {author} {\bibfnamefont {C.-K.}\ \bibnamefont
  {Chiu}}, \bibinfo {author} {\bibfnamefont {J.~C.~Y.}\ \bibnamefont {Teo}},
  \bibinfo {author} {\bibfnamefont {A.~P.}\ \bibnamefont {Schnyder}}, \ and\
  \bibinfo {author} {\bibfnamefont {S.}~\bibnamefont {Ryu}},\ }\href {\doibase
  10.1103/RevModPhys.88.035005} {\bibfield  {journal} {\bibinfo  {journal}
  {Rev. Mod. Phys.}\ }\textbf {\bibinfo {volume} {88}},\ \bibinfo {pages}
  {035005} (\bibinfo {year} {2016})}\BibitemShut {NoStop}%
\bibitem [{\citenamefont {Kitaev}(2006)}]{kitaev2006}%
  \BibitemOpen
  \bibfield  {author} {\bibinfo {author} {\bibfnamefont {A.}~\bibnamefont
  {Kitaev}},\ }\href {\doibase https://doi.org/10.1016/j.aop.2005.10.005}
  {\bibfield  {journal} {\bibinfo  {journal} {Annals of Physics}\ }\textbf
  {\bibinfo {volume} {321}},\ \bibinfo {pages} {2 } (\bibinfo {year} {2006})},\
  \bibinfo {note} {january Special Issue}\BibitemShut {NoStop}%
\bibitem [{\citenamefont {Su}\ \emph {et~al.}(1979)\citenamefont {Su},
  \citenamefont {Schrieffer},\ and\ \citenamefont {Heeger}}]{su1979}%
  \BibitemOpen
  \bibfield  {author} {\bibinfo {author} {\bibfnamefont {W.~P.}\ \bibnamefont
  {Su}}, \bibinfo {author} {\bibfnamefont {J.~R.}\ \bibnamefont {Schrieffer}},
  \ and\ \bibinfo {author} {\bibfnamefont {A.~J.}\ \bibnamefont {Heeger}},\
  }\href {\doibase 10.1103/PhysRevLett.42.1698} {\bibfield  {journal} {\bibinfo
   {journal} {Phys. Rev. Lett.}\ }\textbf {\bibinfo {volume} {42}},\ \bibinfo
  {pages} {1698} (\bibinfo {year} {1979})}\BibitemShut {NoStop}%
\end{thebibliography}%
\bibliographystyle{apsrev4-1}

\newpage
\afterpage{\blankpage}

\appendix

\setcounter{figure}{0}
\makeatletter 
\renewcommand{\thefigure}{S\arabic{figure}}

\newcounter{defcounter}
\setcounter{defcounter}{0}

\newenvironment{myequation}
{%
	\addtocounter{equation}{-1}
	\refstepcounter{defcounter}
	\renewcommand\theequation{S\thedefcounter}
	\align
}
{%
	\endalign
}

\begin{widetext}
	\begin{center}
		{\fontsize{12}{12}\selectfont
			\textbf{Supplemental Material for ``\papertitle''\\[5mm]}}
		{\normalsize \authornames\\[1mm]}
		{\fontsize{9}{9}\selectfont  
			\textit{\tcm}}
	\end{center}
	\normalsize
\end{widetext}

\pagenumbering{gobble}

\subsection{Dissipative SSH Chain}

\textit{Dissipative SSH.---} In the main text, we found that adding symmetry-preserving dissipators to the Kitaev chain will ensure that edge modes of the Lindbladian remain gapless in energy. In closed quadratic superconductors, particle-hole symmetry is generic due to the inherent structure of the fermionic Bogoliubov-de Gennes equation, which leads to a $\mathbb{Z}_2$ classification in the absence of other symmetries in 1D. Thus a single Majorana mode at the edge is protected against all forms of quadratic disorder. Similarly, the spectral matrix $Z$ always satisfies $Z=-Z^*$ (to ensure Hermiticity of the density matrix), implying that a single edge mode will remain gapless in energy upon addition of dissipators which keep the Lindbladian quadratic.

In this section, we provide an example where  gapless edge modes of a closed system can become gapped in energy if dissipators break symmetries of the Hamiltonian. To this end, we turn to dissipative extensions of the Su-Schrieffer-Heeger (SSH) model. Each edge mode of the closed model is composed of two Majorana fermions (one complex fermion). The two Majoranas  can either remain pinned to zero energy, or gap each other out, depending on whether dissipators preserve or destroy the BDI classification of the Hamiltonian. 

Our starting point is the SSH Hamiltonian \cite{su1979}
\begin{myequation} \label{eq:ssh}
\mathcal{H}_{\text{SSH}}= v \sum_i (a_{i,A}^\dagger a_{i,B} + h.c.) + w \sum_i (a_{i,B}^\dagger a_{i+1,A} + h.c.)
\end{myequation}
where $a_{i,A/B}$ annihilates a complex fermion on site $i$ of $N$ in sublattice $A/B$, and $v,w \in \mathbb{R}$. The first-quantized Hamiltonian (matrix) possesses the symmetries: $H_{a}=H_{a}^*, H_{a}=-\Sigma_z H_{a}^* \Sigma_z, H_{a}=-\Sigma_z H_{a} \Sigma_z$ where $\Sigma_z=\mathbb{I}_N \otimes \sigma_z$ and $\sigma_z$ represents the Pauli matrix which acts on the sublattice label within each unit cell. The closed model belongs to class BDI which respects TRS, PHS, and chiral symmetry. We shall now consider two different dissipative scenarios: (1) the case when dissipators respect all three symmetries; (2) the case when dissipators only preserve PHS, resulting in a $\mathbb{Z}_2$ classification, and thereby allowing  two  modes per side of the chain to gap in energy.

\begin{figure}
	\begin{centering}
		\includegraphics[scale=0.12]{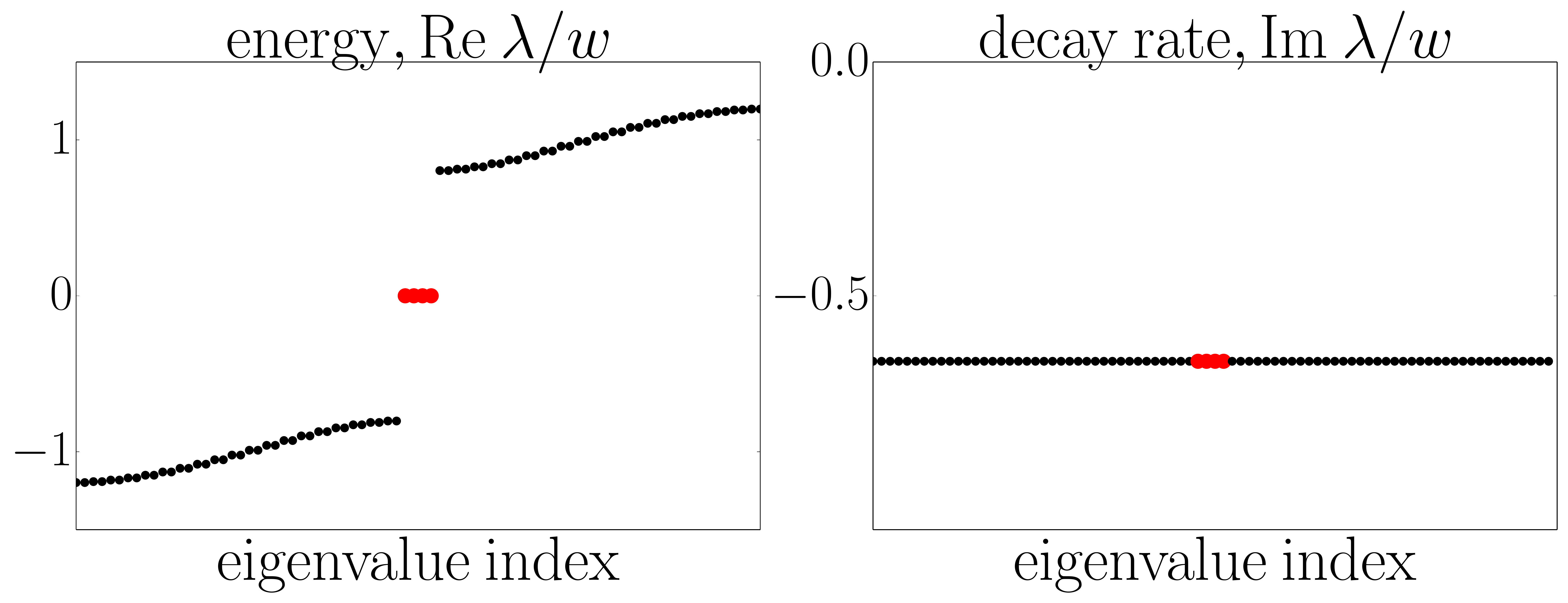}
		\par\end{centering}
	\caption{\label{fig:sshSpecSym}  Lindblad spectrum  for the  SSH model  with symmetry-preserving dissipators \eqref{eq:diss1}, and parameters: $v/w=0.2, \gamma^2/w=0.64$.  The linear dissipators preserve the $\mathbb{Z}$ classification of the closed limit, hence edge modes remain gapless in energy. All modes acquire a uniform decay rate of $\gamma^2/w$.}
\end{figure}

For the symmetry-preserving case, we consider two dissipators per lattice site
\begin{myequation} \label{eq:diss1}
L_{i,A} = \gamma a_{i,A}, \qquad L_{i,B} =  \gamma a_{i,B}^\dagger
\end{myequation}
representing loss on $A$ and gain on $B$ with equal strength. A similar $PT$-symmetric model was considered in Ref.~\cite{dangel2018}. In order to find the spectral matrix $Z$, we must rewrite the Hamiltonian and dissipators in terms of Majoranas: $a_{i,A/B} = \alpha_{i,A/B,a} + \iu  \alpha_{i,A/B,b}$. The first-quantized Hamiltonian matrix splits up into two copies of the Kitaev chain Hamiltonian in a Majorana basis, with non-zero coupling only between different flavors of Majoranas ($a/b$) which belong to different sublattice sites ($A/B$)
\begin{myequation} \label{eq:sshMaj}
H_{\alpha}=\left(\begin{array}{cc}
H_{\text{Kit.}} & 0\\
0 & -H_{\text{Kit.}}
\end{array}\right).
\end{myequation}
$H_{\text{Kit.}}$ is the first-quantized Hamiltonian of Eq.~\eqref{eq:kitchain} in the main text, with hopping magnitudes $v,w$. It satisfies the relations: $H_{\text{Kit.}} = -H_{\text{Kit.}}^*, H_{\text{Kit.}} = \tau_z H_{\text{Kit.}}^* \tau_z, H_{\text{Kit.}} = - \tau_z H_{\text{Kit.}} \tau_z$ where $\tau_z=\mathbb{I}_N \otimes \sigma_z$, and $\sigma_z$ acts on different flavors of Majoranas on opposite sublattice sites. The dissipation matrix in the same basis takes the form: $\iu \text{Re}[M] = \iu \gamma^2  \mathbb{I}_{4N}$, which ensures that $Z$ remains block diagonal and retains  BDI symmetries in each sector.  This type of dissipation preserves the $\mathbb{Z}$ classification of the closed SSH model and hence we expect an arbitrary number of edge modes to remain gapless in energy, which is numerically confirmed in Fig.~\ref{fig:sshSpecSym}.

\begin{figure}
	\begin{centering}
		\includegraphics[scale=0.12]{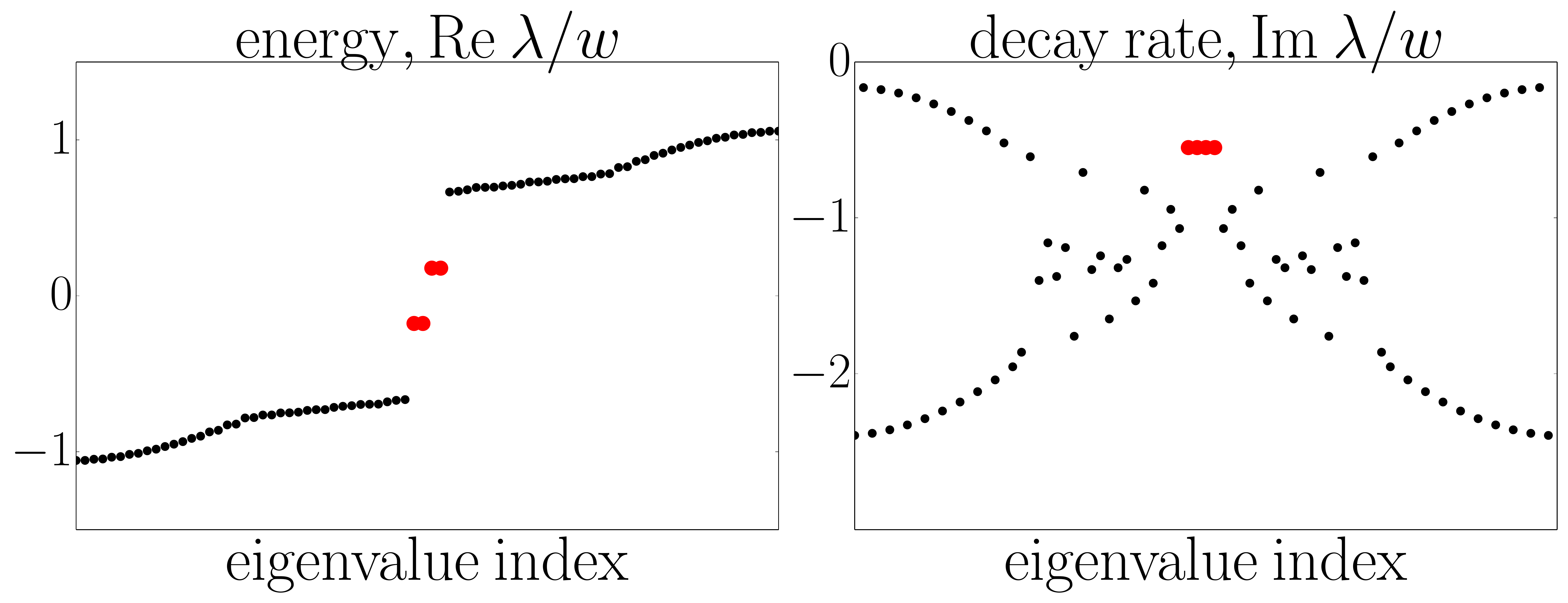}
		\par\end{centering}
	\caption{\label{fig:sshSpecBreak}  Lindblad spectrum for the SSH model with symmetry-breaking dissipators  \eqref{eq:diss2}, and parameters: $v/w=0.2, \gamma^2/w=0.64, \theta=\pi/4$.  The linear dissipators break pseudo-anti-Hermiticity of the quadratic Lindbladian, hence two edge modes per side of the chain (red dots) can start to gap in energy. While the closed SSH model belongs to class BDI, dissipators evolve it to the trivial $\mathbb{Z}_2$ sector of non-Hermitian class D.}
\end{figure}

For the symmetry-breaking case, we  choose $2N-1$ nearest-neighbor dissipators
\begin{myequation} \label{eq:diss2}
L_{j,A} = \gamma(a_{j,A} + e^{\iu \theta} a_{j,B}),\qquad L_{j,B} = \gamma(a_{j,B} + e^{\iu \theta} a_{j+1,A}).
\end{myequation}
 The Hamiltonian in a Majorana basis is the same as in \eqref{eq:sshMaj}, but now the dissipation matrix begins to couple the two blocks. The resulting matrix: $Z=H_\alpha + \iu \text{Re}[M]$ only possesses the symmetry: $Z=-Z^*,$ i.e. all its elements are purely imaginary.  This implies that the model falls into class D with  a $\mathbb{Z}_2$ classification. Moreover, since each edge mode is composed of two Majorana fermions in the closed limit, this type of dissipation forces the model into the trivial  sector of $\mathbb{Z}_2$. We therefore expect fragile edge modes in this dissipative extension of  SSH which is numerically confirmed   in Fig.~\ref{fig:sshSpecBreak}. The energy splitting of the edge modes scales as: $(\gamma^2/w)^2$ at small $\gamma$, hence we need relatively strong values of dissipation for this effect to be noticeable. Nevertheless, we have demonstrated that linear dissipation can break symmetries of the closed model, resulting in gapless edge modes which are fragile against certain dissipative channels.

\subsection{Time-reversal symmetry in Lindbladians}

Here, we discuss a physical interpretation of the symmetry \eqref{eq:trs} which reduces to the usual Hermitian time-reversal symmetry in the dissipation-free limit. We will show that this system naturally arises when the Lindbladian describes a system whose microscopic Hamiltonian for the system and environment respects the Hermitian time-reversal symmetry \eqref{eq:trsHam}. In doing so, we follow the derivation of the Lindblad master equation found in Ref.~\cite{breuer2002} (Section 3.3).

The starting point for this derivation is a (Hermitian) Hamiltonian for the combined system and environment, which together form an isolated system:
\begin{myequation}
\hat{H} = \hat{H}_S + \hat{H}_B + \hat{H}_{SB}
\label{eqHamTotal}
\end{myequation}
where $\hat{H}_S$ acts on the system, $\hat{H}_B$ acts on the bath, and $\hat{H}_{SB}$ couples the two, and is assumed to be weak. The system-bath coupling can always be decomposed into $m$ channels according to
\begin{myequation}
\hat{H}_{SB} = \sum_{\alpha = 1}^m \hat{A}_\alpha \otimes \hat{B}_\alpha,
\label{eq:intdecomp}
\end{myequation}
where $\hat{A}_\alpha$ and $\hat{B}_\alpha$ are operators acting on the system and bath respectively. One can always demand that they are separately Hermitian $\hat{A}_\alpha = \hat{A}_\alpha^\dagger$, $\hat{B}_\alpha = \hat{B}_\alpha^\dagger$. The $\hat{A}_\alpha$ operators can be further decomposed in the energy eigenbasis of $\hat{H}_S$. Specifically, $\hat{A}_\alpha(\omega)$ is defined as the component of $\hat{A}_\alpha$ which excites the system by an energy $\omega$:
\begin{myequation}
\hat{A}_\alpha(\omega) = \sum_{\epsilon' - \epsilon = \omega} \hat{P}(\epsilon) \hat{A} \hat{P}(\epsilon')
\end{myequation}
where $\hat{P}(\epsilon)$ is the  projector onto the eigenspace of $\hat{H}_S$ with energy $\epsilon$.

In the weak coupling regime, wherein the Born approximation can be applied, the system and bath can be described by a factorized density matrix $\hat{\rho}_S(t) \otimes \hat{\rho}_B$, where the bath density matrix is assumed to be stationary $[\hat{\rho}_B, \hat{H}_B] = 0$. The bath correlation functions can then be defined as
\begin{myequation}
	\Gamma_{\alpha \beta}(t) \coloneqq \Braket{\hat{B}_\alpha (t)^\dagger \hat{B}_\beta(0)} \equiv \text{Tr} \left(\hat{B}_\alpha (t)^\dagger \hat{B}_\beta(0) \hat{\rho}_B \right),
\end{myequation}
where the time evolution is calculated using $\hat{H}_B$ only. This in turn gives the bath spectral functions
\begin{myequation}
	\gamma_{\alpha \beta}(\omega) = \int_{-\infty}^{+\infty} {\rm d} t\, e^{\iu \omega t} \Gamma_{\alpha \beta}(t).
\end{myequation}
With these quantities defined, a standard derivation yields a non-diagonal form for the Lindblad master equation \cite{lindblad1976,breuer2002}
\begin{myequation}
\mathcal{L}\hat{\rho} &= [\hat{H}_S + \hat{H}_{LS}, \hat{\rho}]  + \iu \sum_{\alpha \beta} \sum_\omega \gamma_{\alpha \beta}(\omega) \left[\hat{A}_\beta(\omega) \hat{\rho}_S \hat{A}_\alpha(\omega)^\dagger \vphantom{\frac{1}{2}}\right. \nonumber\\ &- \left. \frac{1}{2} \left\{\hat{A}_\alpha(\omega)^\dagger \hat{A}_\beta(\omega), \hat{\rho}_S \right\} \right],
\end{myequation}
where $\hat{H}_{LS}$ is the Lamb shift -- a renormalization of the system Hamiltonian by the action of the bath. The above can be written in the standard form \eqref{eq:lindMaster} by diagonalizing $\gamma_{\alpha \beta}(\omega)$ at each omega.

We now insist that the microscopic Hamiltonian \eqref{eqHamTotal} respects a (second quantized) time-reversal symmetry $\hat{U}_S \hat{U}_B \hat{H}^* \hat{U}_B^\dagger \hat{U}_S^\dagger = \hat{H}$. Here, $\hat{U}_S$ and $\hat{U}_B$ are unitary matrices acting on the system and bath degrees of freedom, respectively. In a fermionic system, these operators satisfy $\hat{U}_S \hat{U}_S^* = (\pm 1)^{\hat{P}_F}$, where $\hat{P}_F$ is the fermion parity operator, and the choice $\pm 1$ correspond to integer ($+1$) and half-integer $(-1)$ spins \cite{chiu2016}. We will find that this symmetry imposes constraints on the resulting form of the Lindbladian, which in quadratic form ensures that the non-hermitian time-reversal symmetry \eqref{eq:trs} is satisfied.

In terms of the decomposition \eqref{eq:intdecomp}, the sufficient and neccesary conditions for $\hat{H}_{SB}$ to be time-reversal symmetric are
\begin{myequation}
	\hat{U}_S \hat{A}^*_\alpha \hat{U}_S^\dagger &= e^{\iu \theta_\alpha} \hat{A}_\alpha \nonumber\\
	\hat{U}_B \hat{B}^*_\alpha \hat{U}_B^\dagger &= e^{-\iu \theta_\alpha} \hat{B}_\alpha
	\label{eqTRSMicro}
\end{myequation}
where $\theta_\alpha$ is a phase, which is constrained to be $0$ or $\pi$ by the Hermiticity of $\hat{A}_\alpha$ and $\hat{B}_\alpha$. Thus $\hat{A}_\alpha$ and $\hat{B}_\alpha$ must be both odd or both even under TRS. Furthermore, given that $\hat{H}_S$ and $\hat{H}_B$ are also time-reversal symmetric, we have the same relations for $\hat{A}_\alpha(\omega)$ and $\hat{B}_\alpha(\omega)$.

When \eqref{eqTRSMicro} is satisfied, the bath correlation function acquires a symmetry
\begin{myequation}
\Gamma_{\alpha \beta}(t) &= \Tr \left(\hat{B}_\alpha (t)^\dagger \hat{B}_\beta(0) \hat{\rho}_B \right) \nonumber\\
&= \Tr \left(\hat{B}_\alpha (t)^T \hat{B}_\beta(0)^* \hat{\rho}_B^* \right)^* \nonumber\\
&= e^{\iu(\theta_\beta - \theta_\alpha)}\Tr \left( \hat{U}_B^\dagger\hat{B}_\alpha (-t)^\dagger \hat{U}_B\hat{U}_B^\dagger \hat{B}_\beta(0) \hat{U}_B\hat{U}_B^\dagger \hat{\rho}_B \hat{U}_B \right)^* \nonumber\\
&= e^{\iu(\theta_\beta - \theta_\alpha)}\Tr \left(\hat{B}_\alpha (-t)^\dagger \hat{B}_\beta(0) \hat{\rho}_B \right)^* \nonumber\\
&= e^{\iu(\theta_\beta - \theta_\alpha)} \Tr \left( \hat{\rho}_B \hat{B}_\beta(0)^\dagger \hat{B}_\alpha (-t) \right) \nonumber\\ &\equiv e^{\iu(\theta_\beta - \theta_\alpha)} \Gamma_{\beta \alpha}(t),
\end{myequation}
which supplements the Hermiticity condition $\Gamma_{\alpha \beta}(t) = \Gamma_{\beta \alpha}(-t)^*$. The spectral function $\gamma_{\alpha \beta}(\omega)$ is then a Hermitian positive semi-definite $m \times m$ matrix constrained by
\begin{myequation}
\Lambda \gamma(\omega)^* \Lambda = \gamma(\omega)
\label{eq:specsym}
\end{myequation}
where $\Lambda$ is a diagonal matrix with elements $e^{\iu \theta_\alpha}$ and we have suppressed the channel indices $\alpha$, $\beta$. At each energy $\omega$, one can define $m$ jump operators $\hat{L}_\mu$ which take the form
\begin{myequation}
	\hat{L}_\mu = \sqrt{\kappa_\mu} \sum_\alpha u_\mu^\alpha \hat{A}_\alpha(\omega),
\end{myequation}
where $u_\mu$ is an eigenvector with components $u_\mu^\alpha$ satisfying $\gamma(\omega) u_\mu = \kappa_\mu u_\mu$, with $\kappa_\mu$ the real, non-negative eigenvalue.  We consider the action of the TRS operation on the jump operators, namely
\begin{myequation}
\hat{U}_S \hat{L}_\mu^* \hat{U}_S^\dagger &= \sqrt{\kappa_\mu} \sum_\alpha (u_\mu^\alpha)^* \hat{U}_S \hat{A}_\alpha(\omega)^* \hat{U}_S^\dagger \nonumber\\
&= \sqrt{\kappa_\mu} \sum_\alpha e^{\iu \theta_\alpha}  (u_\mu^\alpha)^* \hat{A}_\alpha(\omega).
\end{myequation}
The symmetry condition on the bath spectral functions \eqref{eq:specsym} implies that the eigenvectors satisfy $\Lambda u_\mu^* = u_\mu$, and so we find that the jump operators must satisfy
\begin{myequation}
	\hat{U}_S \hat{L}_\mu^* \hat{U}_S^\dagger  = \hat{L}_\mu.
\end{myequation}
Now, the jump operators are assumed to be linear in the fermionic operators according to \eqref{eq:quadratic}. The action of TRS on the Majorana operators determines the first quatized symmetry operator $U_T$ (which is a $2N \times 2N$ matrix) via $\hat{U}_S \hat{\alpha}_j^* \hat{U}_S^\dagger = \sum_{k=1}^{2N} (U_T)_{jk} \hat{\alpha}_k$ \cite{chiu2016}, and so we find that the coefficients $l_{\mu, j}$ must satisfy
\begin{myequation}
\sum_k (U_T)_{jk} l_{\mu, k}^* = l_{\mu, j}.
\label{eq:trsJump}
\end{myequation}
In a similar way, one can also verify that the Lamb shift $\hat{H}_{LS}$ induces a quadratic correction to the system Hamiltonian, such that the Hamiltonian part of the Lindbladian respects the first quantized Hermitian TRS \eqref{eq:trsHam}. Combining these results, using the definition $Z = H + \iu \Re M$, one finds that $U_T Z^T U_T^\dagger = Z$, as desired.

We therefore see that the non-Hermitian time-reversal symmetry \eqref{eq:trs} is satisfied in systems where the microscopic Hamiltonian for the system and bath, which is itself isolated, respects a Hermitian TRS. Note, however, that the system still propagates irreversably. The above is a sufficient but not necessary condition for \eqref{eq:trs} to be satisfied. This is because $Z$ is independent of the imaginary part of $M$, which does not affect the spectrum of the Lindbladian. One could in principle construct a system in which $Z$ satisfies the non-Hermitian TRS, but the imaginary part of $M$ is chosen such that the condition \eqref{eq:trsJump} is violated. However, we expect that such a scenario would require fine-tuning, and therefore does not represent a generic symmetry condition.
\\

\subsection{Independence of spectral and steady-state properties}

In this section, we demonstrate that the robust spectral features of quadratic Lindbladians discussed in the main text are independent of properties of the steady state, the latter of which was the subject of study in Refs.~\cite{bardyn2013, bardyn2018, budich2015b, viyuela2014}. Specifically, we show that any system with some topological features in its spectrum can be continuously deformed to a system with the same spectrum, but a trivial (infinite temperature) steady state, while at all times maintaining the relevant spectral gaps, symmetries, and locality of the equations of motion.

In a quadratic system, the steady state density matrix can be completely characterized by its two point correlation functions
\begin{myequation}
	\Gamma_{ij} \coloneqq \frac{1}{4i}\left[\Tr\left(\hat{\alpha}_i \hat{\alpha}_j \hat{\rho}_{SS}\right) - \delta_{ij}\right].
\end{myequation}
With the Lindbladian written in the form \eqref{eq:Lindblad1}, this correlation matrix is determined by the Sylvester equation \cite{prosen2010}
\begin{myequation}
Z^T \Gamma + \Gamma Z =  \iu \Im[M]
\label{eqSylvester}
\end{myequation}
where $Z=H+\iu \Re M$.  We assume that all eigenvalues of $Z$ have a non-zero imaginary part, which implies that the solution to \eqref{eqSylvester} is unique, and that the  topological properties of the steady state are well-defined \cite{bardyn2013}.

Our aim is to construct a continuous path of quadratic Lindbladians parametrized by $s \in [0,1]$ which interpolates between the physical system at $s=0$, and a system with the same spectrum (and therefore the same robust spectral features), but a trivial steady state at $s=1$. Because the space of physical generators $Z$ obeys a complicated set of constraints which enforce positivity, we choose to define this path at the level of the Hamiltonian $H_{i,j}$ and the jump operators $l_{\mu, i}$, which are constrained only by the relevant symmetries \eqref{eq:TFWZ}.
%We specify this path at the level of the Hamiltonian $H_{i,j}$ and the jump operators $l_{\mu, i}$, which are constrained only by the relevant symmetries \eqref{eq:TFWZ}, whereas the space of physical generators $Z$ obeys a  more complicated set of constraints, in order to ensure that $\Im [\lambda_i] \leq 0$.

We find it useful to separate out the real and imaginary parts of $l_{\mu, i}$ into two independent matrices as
\begin{myequation}
(A)_{\mu i} &= \Re [\ell_{\mu, i}], & (B)_{\mu i} &= \Im [\ell_{\mu, i} ].
\end{myequation}
Without loss of generality, we take the number of independent channels $\{\mu\}$ to be $2N$, since any linearly dependent set of jump operators can be reduced to a linearly independent set without changing the equations of motion \cite{breuer2002}. We then have $A$ and $B$ square, with
\begin{myequation}
	Z = H + \frac{\iu}{2} \left(A^TA + B^TB\right).
\end{myequation}

Our strategy is to deform the system by adiabatically turning off $B$, whilst adjusting $A$ so that $Z$ (which determines the spectrum of the Lindbladian) is constant throughout. At the end of the evolution, we will have $\Im[M] = A^TB - B^TA = 0$, such that the unique solution to the steady state equation \eqref{eqSylvester} is $\Gamma = 0$, which corresponds to a trivial infinite temperature state $\hat{\rho} \propto \mathbbm{1}$. %As previously mentioned, we wish to specify the deformation procedure at the level of the unconstrained quatities $H_{ij}$ and $\ell_{\mu, i}$ (Equivalently, $H, A, B$).
We start by defining $B(s)$ throughout the evolution as
\begin{myequation}
B(s) = (1-s)B
\end{myequation}
which interpolates between $B(s=0) = B$ and $B(s=1) = 0$. Now we adjust $A$ to compensate in a way that ensures $Z(s) = H(s) + \iu (A(s)^TA(s) + B(s)^TB(s))/2$ is constant and equal to the physical $Z(s) = Z(s=0) = Z$. This means that $H$ can remain independent of $s$, and $A(s)$ must satisfy the equation
\begin{myequation}
A(s)^TA(s) &= A^TA + B^TB - B^TB(1-s)^2 \nonumber\\
&= A^TA + [s(2-s)]B^TB.
\label{eqACondition}
\end{myequation}

Now $A^TA$ and $B^TB$ are necessarily real, symmetric positive semi-definite matrices, and the factor $s(2-s)$ is non-negative for $0 \leq s \leq 1$. This means that the right-hand side of \eqref{eqACondition} is also a real, symmetric positive semi-definite matrix, and thus can indeed be represented in the form $A(s)^T A(s)$. The solution $A(s)$ is not unique, since it can be left multiplied with an arbitrary orthogonal matrix.

In order for topological properties to be robust, the dissipators $\hat{L}_\mu$ must be local throughout the deformation. This implies that the solution $A(s)$ must be chosen such that its rows have components which are spatially localized. Specifically, for all $\mu$, $A(s)_{\mu j}$ must decay sufficiently quickly with $j$ away from some site $j_\mu$ (strictly faster than $|j-j_\mu|^{-d}$ in spatial dimension $d$ \cite{kitaev2006}). The same locality condition holds at $s=0$ for the physical system, and thus $(A^TA)_{jk}$ and $(B^TB)_{jk}$ have components that decay equally quickly with $|j-k|$. This means that a sufficiently local solution $A(s)$ of Eq.~\eqref{eqACondition} can always be found. For example, one can verify that the standard Choleskey decomposition of \eqref{eqACondition} gives a solution $A(s)$ with the same locality properties as $A$. From this local upper-diagonal solution, $A(s)$ can be further rotated by a locality-preserving orthogonal matrix $A(s) \rightarrow Q(s)A(s)$ ($Q(s)^TQ(s) = 1$), chosen such that the deformation is contiunous in $s$.

With this form of $A(s)$ and $B(s)$, we choose a constant Hamiltonian $H(s) = H$. Together, this ensures that $Z(s)$ remains independent of $s$ throughout the evolution, so that the symmetries and spectral properties of the Lindbladian are preserved throughout, while the steady state gradually evolves into one with a covariance matrix $\Gamma = 0$.

In conclusion, we have defined a deformation procedure which preserves the necessary symmetry and locality properties, and interpolates between the physical system and one with $B = 0$, whilst keeping the matrix $Z$ (and consequently its spectrum) constant throughout. At the end of the deformation $s = 1$, we have $\Im[M] = 0$ and therefore the only solution to the steady-state equation \eqref{eqSylvester} is $\Gamma = 0$ (the infinite temperature state $\hat{\rho}_{SS} \propto \mathbbm{1}$) which is trivial. We conclude that it is always possible to interpolate between the physical steady state and a trivial one without changing any of the features of the spectrum. Therefore topological edge modes in the spectrum can be supported without any topological features in the steady state.

\end{document}